\documentclass[aip, jap,
amsmath,amssymb,
reprint, 
]{revtex4-1}

\usepackage[colorlinks=true, linkcolor=blue, citecolor=blue, urlcolor=blue]{hyperref}
\usepackage{xcolor}
\usepackage{graphicx}
\usepackage{dcolumn}
\usepackage{bm}
\usepackage[english]{babel}
\usepackage{orcidlink}

\usepackage{caption}          
\usepackage{subcaption}
\usepackage{algorithm}
\usepackage{tabularx}  
\usepackage{afterpage}
\usepackage{color,soul}
\usepackage{lmodern}
\usepackage{float}

\makeatletter
\def\@email#1#2{%
 \endgroup
 \patchcmd{\titleblock@produce}
  {\frontmatter@RRAPformat}
  {\frontmatter@RRAPformat{\produce@RRAP{*#1\href{mailto:#2}{#2}}}\frontmatter@RRAPformat}
  {}{}
}%
\makeatother

\begin{document}

\preprint{AIP/123-QED}

\title[]{Influence of Turbulence Length Scale and Platform Surge Motion on Wake Dynamics in Tandem Floating Wind Turbines}


\author{Ahmad Nabhani\orcidlink{0009-0008-9828-1264}} 
\affiliation{Fluid Mechanics Department, Universitat Politècnica de Catalunya, 08034 Barcelona, Spain}
\affiliation{YPlasma Actuator Technologies Inc., Research and Development Division, 08039 Barcelona, Spain}


\author{Josep M. Bergada \orcidlink{0000-0003-1787-7960}}
\affiliation{Fluid Mechanics Department, Universitat Politècnica de Catalunya, 08034 Barcelona, Spain}
\email[Correspondance: ]{josep.m.bergada@upc.edu}

\date{\today}

\begin{abstract}
Wake interaction is a key factor limiting the performance of floating offshore wind turbine arrays, yet the combined influence of inflow turbulence structure and platform motion on wake dynamics remains poorly understood. This study examines how the integral length scale of inflow turbulence and platform surge motion shapes wake development and power performance in a tandem configuration of two aligned floating offshore wind turbines separated by five rotor diameters.
High-fidelity computational fluid dynamics simulations are performed using OpenFOAM, based on Large-Eddy Simulation with an Actuator-Line Model and the Wall-Adapting Local Eddy-Viscosity (WALE) subgrid-scale closure. Synthetic turbulent inflows are generated using the Divergence-Free Synthetic Eddy Method, with prescribed integral length scales spanning $0.25$–$1.25$ times the rotor radius. Over this range, increasing the integral length scale naturally leads to higher freestream turbulence intensity, which increases from approximately $1.9\%$ to $7.2\%$. The corresponding dominant inflow frequencies are extracted from time-resolved velocity signals, yielding Strouhal numbers in the range $St \approx 0.71$ to $0.12$. Wake evolution is analyzed through disk-averaged velocity deficits, turbulent kinetic energy distributions, spectral characteristics, vortex topology, and time-averaged power coefficients.
The results show that the inflow turbulence integral length scale is the primary parameter controlling wake recovery. Larger integral scales introduce energetic, low-frequency eddies that destabilize the tip-vortex system, enhance lateral and vertical entrainment, and accelerate wake mixing. These mechanisms lead to substantial reductions in the inter-turbine velocity deficit and translate directly into increased downstream power output. Depending on the turbine configuration, downstream power gains range from approximately $+90\%$ to $+140\%$ relative to uniform inflow conditions. Platform surge motion of the upstream turbine further promotes wake unsteadiness and contributes to additional power recovery, whereas the relative phase of surge motion between turbines has only a minor effect on the mean wake and time-averaged power.
These findings demonstrate that downstream performance in floating wind turbine arrays is governed predominantly by upstream wake dynamics and the temporal and spatial structure of the incoming turbulence. The study provides a physics-based understanding of turbulence–motion coupling in floating turbine wakes and highlights the important role of 
\textcolor{black}{turbulence scales/turbulence intensity} when assessing wake interactions in realistic offshore environments.

\end{abstract}

\maketitle

\section{\label{sec:itro} Introduction}

Wind energy has developed rapidly in recent years amid energy crises and, more importantly, due to the effectively inexhaustible nature of the resource and the mature technology of wind turbines \cite{REN2018380}. It is expected that wind energy will supply between one-quarter and one-third of global electricity demand by 2050 \cite{Veers2019}. To meet this demand, more wind farms need to be installed both onshore and offshore. The offshore environment is an ideal site for wind farms, as it offers abundant wind energy resources and economic advantages such as higher capacity factors, and also high and stable utilization of grid assets \cite{SATYMOV2025125980}.

At sea, floating offshore wind turbines (FOWTs) have been projected as a more favorable approach for offshore wind energy harvesting than their bottom-fixed counterparts \cite{Veers2019, james2015floating, wes-1-1-2016}, but many designs are still at the prototype stage with limited performance data \cite{BARTER20201}. In a floating wind farm, the majority of turbines operate in the wake, which refers to the airflow formed after the wind passes through a turbine, leading to a region of reduced wind speed \textcolor{black}{and a considerable velocity deficit, resulting in a loss of over 10\% of annual energy production \cite{Barthelmie2009} and increasing fatigue loads on downstream turbines \cite{troldborg2011numerical, VERMEER2003467, MENG2020788}.} In addition, real wind turbines operate in the lowest part of the atmospheric boundary layer (ABL), where the flow conditions are inherently unsteady and fluctuate across a wide range of length and time scales, which affect power generation efficiency and wake dynamics in real wind farms \cite{porte2020wind, Wei_El}. In addition to turbulence intensity, recent large-eddy simulation (LES) studies have shown that surface-layer stability and free-atmosphere stratification strongly influence wake recovery, turbulence levels, and wind farm power output, with stable conditions leading to slower recovery, enhanced blockage, and gravity-wave generation \cite{Souaiby2025Atmospheric}.

Several factors and mechanisms, including turbine layout and spacing \cite{andersen2017turbulence, hodgson2025impact}, platform motion of floating wind turbines \cite{li2022onset, Messmer_Hölling_Peinke_2024}, and dynamic flow control strategies \cite{frederik2020helix, munters2018towards}, introduce motions spanning a wide range of scales at both the turbine and farm levels. Insights into the combined effects of ABL turbulent flow and the scale motions introduced by the aforementioned factors and mechanisms can potentially inspire new strategies for the design, optimization, and analysis of floating wind farms, for more efficient, reliable, and safe wind energy production \cite{dabiri2020theoretical}. 
However, the interaction between the multiscale turbulence of the atmospheric flow and the unsteady motions of floating wind turbines introduces additional complexity to wake entrainment and recovery processes. Platform motions generate time-dependent disturbances that interact with the turbulence-induced shear layers and coherent wake structures, modifying the transfer of momentum and energy within and around the wake. Despite its importance for floating wind turbine arrays, this coupled turbulence–motion interaction has received extremely limited investigation to date \cite{Veers2019, meneveau2019big, porte2020wind, vahidi2024influence, hodgson2025impact}.

The influence of inflow turbulence intensity on wake recovery of a wind turbine is relatively well explored: considering a given surface roughness and a stable ABL, higher inflow turbulence intensity results in faster wake recovery \cite{Churchfield01012012, barlas2016roughness, en15082899}, due to the higher momentum interchange between particles associated with the incoming fluid flow and the wake \cite{Churchfield01012012, en5125340}. For instance, Carbajo Fuertes et al.\cite{rs10050668}, and Gambuzza et al.\ \cite{gambuzza2023influence} highlighted a linear relation between freestream turbulence intensity and wake recovery rate. Inflow turbulence also influences the evolution of coherent flow structures in wind turbine wakes. Both atmospheric flows and wind turbine wakes contain flow structures over a wide range of scales \cite{PhysRevFluids.1.063701, porte2020wind}, and their interaction influences wake evolution in many ways, including: (i) formation and breakdown of tip vortices, which are typical flow structures in the near-wake region and separate the low-momentum wake flow from the high-momentum ambient flow; and (ii) low-frequency, large-scale lateral motion of the far wake, referred to as wake meandering\cite{sorensen2015simulation, li2024impacts}.  

Far-wake meandering impacts both the wake recovery and the loading of downstream turbines, and has been studied through a combination of numerical simulations (mostly using large eddy simulation), experimental studies with small-scale model turbines, and field measurements \cite{hogstrom1988field, espana2011spatial, larsen2007dynamic, yang2016coherent}. This phenomenon is attributed to both the influence of large atmospheric scales \cite{larsen2007dynamic} and bluff-body-type vortex shedding via shear-layer or tip-vortex instability \cite{posa2021instability, biswas2024effect, larsen2008wake}, and can offer useful insights into the generation of wake meandering. When considering the dynamics of wake meandering \cite{larsen2008wake, madsen2010calibration}, 
the turbine wake is thought to be passively advected by lateral and vertical atmospheric scales with a length scale larger than $2D$ (where $D$ is the rotor diameter) \cite{hodgson2023effects, li2024impacts}. Inflow perturbations with frequencies in the range $0.25 < St < 0.3$ (where $St = fD/U_\infty$ is the Strouhal number) are particularly effective at triggering wake meandering \cite{mao2018far, gupta2019low}.

Recently, a new mechanism has been proposed suggesting that wind turbine wakes are not simply advected by the flow structures of the freestream, as assumed in the classical dynamic meandering model \cite{larsen2008wake}. Instead, the turbine exhibits a dynamic response. That is, the wake interacts with specific inflow turbulence scales (specific freestream turbulence frequencies), and these scales preferentially excite wake meandering \cite{li2024impacts, gupta2019low, heisel2018spectral, hodgson2025impact}. 
Focusing on the dynamics of individual turbine wakes, Mao and Sørensen \cite{mao2018far} studied wake meandering and identified non-linear optimal perturbations to a wind turbine wake modeled with an actuator disk model (ADM). They found that at a Reynolds number of $Re = 1000$ (based on the free-stream velocity and turbine radius), the most energetic inflow perturbation is a large-scale, low-frequency azimuthal mode with wavenumber $m = 1$ and a characteristic Strouhal number in the range $St \approx 0.08$--$0.2$. As the Reynolds number increases, the dominant wake oscillation frequency (meandering frequency) decreases and converges to $St \approx 0.16$ at $Re = 3000$, remaining approximately constant at higher Reynolds numbers. These low frequencies fall well within the accepted range of wake meandering frequencies \cite{chamorro2013interaction, okulov2014regular}.

Similarly, oscillations in the far-wake region may result from the selective amplification of upstream flow structures, which was also observed by Sarmast et al.\cite{Sarmast_2014} in an LES study. However, for $Re > 3000$, it was observed that the wake oscillation frequency did not change, regardless of the Reynolds number of the incoming flow. Such minimal impact of Reynolds number differences was also reported by McTavish et al.\cite{mctavish2013evaluating}. In field measurements, Heisel et al.\cite{heisel2018spectral} also showed that the wind turbine amplifies a range of inflow frequencies, yielding spectral peaks in the far wake in the range of $St = 0.3$--$0.4$, consistent with the meandering band noted above. 
Larger freestream characteristic length scales were shown to induce stronger wake meandering and were linked to faster kinetic energy (KE) entrainment from the freestream flow to the wake region \cite{vahidi2024influence, blackmore2014influence}. It is worth noting that these studies modeled a turbine using an actuator disk model, so the effects of tip vortices were not accurately considered. The results of these studies likely hold better in the far wake of a turbine, where the wind turbine wake is better approximated by the wake of an actuator disk \cite{biswas2025effect}.

Recent studies have demonstrated that wake recovery is strongly influenced by the characteristic scales of inflow turbulence, and not by turbulence intensity alone. 
Li et al. \cite{li2024impacts} investigated these effects using large-eddy simulations of a permeable rotor disk model under inflows with a wide range of turbulence intensities ($I_u \approx 2.5\%$–$25\%$) and integral length scales ($L_u/D \approx 0.5$–$2.0$). They showed that both increased turbulence intensity and increased integral length scale promote wake recovery by enhancing ejection and sweep events at the wake boundary. Turbulence intensity primarily determined where wake recovery begins, while the integral length scale mainly controlled the recovery rate downstream. Importantly, they showed that integral length scales larger than $0.5D$ particularly increased the extraction of turbulent kinetic energy (TKE) by the turbine. They highlighted that turbulence scales influence wake recovery through both energetic and spectral mechanisms.

Using LES under natural atmospheric conditions, Vahidi and Porte-Agél \cite{vahidi2024influence} studied the effect of inflow turbulent length scales by varying the boundary-layer height. They found that larger inflow length scales led to faster wake recovery, mainly due to enhanced turbulent momentum fluxes and stronger entrainment. At larger scales, wake meandering was more pronounced in both the lateral and vertical directions. 

Experimental evidence from Biswas and Buxton \cite{biswas2025effect} further clarified the role of turbulent scales in wake recovery. Using particle image velocimetry, they studied a model wind turbine subjected to free-stream turbulence intensities ($TI_\infty$) and length scales ($L_u$) in the range $1\% < TI_\infty < 13.1\%$ and $0.01D < L_u < 0.11D$, respectively. Increasing either turbulence intensity or integral length scale caused earlier breakdown of tip vortices, which in turn led to an earlier onset of wake recovery. Wake meandering became stronger in the presence of free-stream turbulence, though not monotonically with either parameter. Their analysis revealed that wake meandering under turbulent inflow more effectively extracted energy from the mean velocity shear, while nonlinear interactions and diffusion also contributed to wake recovery.

Bourhis et al. \cite{bourhis2025impact} extended this understanding by jointly varying turbulence intensity, integral length scale, and thrust coefficient ($C_T$). For $1\% \lesssim TI_\infty \lesssim 11\%$ and $0.1 \lesssim L_u/D \lesssim 2$, they identified distinct near- and far-wake regimes. In the near wake, higher turbulence intensity and higher thrust coefficient promoted faster wake growth. In contrast, in the far wake, increasing turbulence intensity reduced the wake growth rate, introducing a turning point in the wake recovery process. They reported that for low-$TI$ and small-$L_u$, wake meandering is minimal and sensitive to $C_T$. \textcolor{black}{Wake meandering is enhanced for high-$TI$ and large-$L_u$, where the integral length scale plays a leading role.}

At even larger scales, Hodgson et al. \cite{hodgson2025impact} investigated the influence of inflow turbulence characteristics using large-eddy simulations with actuator line modeling. When considering the wind-farm configuration, the inflow turbulence integral length scale was varied over the range
$L_u \in [3.2R,\;12.0R]$, while maintaining the mean inflow velocity. 
They demonstrated that shorter inflow integral length scales resulted in a faster breakdown of the near wake behind the first row of turbines and enhanced wake recovery compared to longer integral length scales. 
Focusing on inflow turbulence integral time scales, experimental work by Gambuzza and Ganapathisubramani \cite{gambuzza2023influence} examined inflows with integral time scales spanning up to $10$ times the convective time scale based on the rotor diameter ($D/U_{\infty}$). They evaluated free-stream turbulence intensities between $3\%$ and $12\%$. 
They observed that high turbulence intensity combined with short integral time scales led to rapid wake evolution and earlier wake recovery. 
In contrast, large integral time scales delayed wake development even under highly turbulent conditions. 
This delayed recovery was attributed to stabilization of the near-wake shear layer and the persistence of coherent helical vortex structures, which suppressed momentum exchange at the wake boundary.

Similarly, in a numerical study, Hodgson et al.\cite{hodgson2023effects} investigated the effect of inflow turbulent time scales by varying the dominant Strouhal number over the range $St \approx 0.04$--$0.5$ under Taylor’s frozen-turbulence hypothesis, while maintaining identical turbulence intensity at the computational domain inlet. 
They found that shorter 
inflow time scales promoted an earlier breakdown of the near wake and faster wake recovery, whereas longer time scales delayed wake evolution. 

When considering the tip-speed ratio (TSR) as an additional parameter, LES studies have shown that as the TSR increases, the onset position of wake meandering moves closer to the rotor and the magnitude of wake oscillations becomes stronger \cite{wang2025effects}. Anderson et al.\ \cite{andersen2017turbulence} demonstrated that the downstream turbine power output strongly depends on both the inflow integral time scale and the turbine TSR (in addition to the turbine location), showing a dynamic interaction between inflow and rotor time scales.

In offshore environments, platform motion adds another unsteady input that can distort near-wake structures and modulate recovery. Prior computational studies have investigated the influence of surge and pitch motions of floating wind turbines and reported distortion of near-wake structures \cite{TRAN2016204, LIU201695, LEE20199}. Wind tunnel measurements have also been used to analyze their influence on the wake dynamics of FOWTs \cite{Fontanella2021UNAFLOW, Fontanella_2022, ROCKEL20171, ROCKEL2016666, Wei2022phase, Wei_Dabiri_2023, messmer2025role}. Fontanella et al.\cite{Fontanella_2022} observed traveling-wave oscillations in the streamwise velocity downstream of a periodically surging wind turbine, but at a distance of $2.3D$ into the wake, no change in wake recovery was found. Rockel et al.\cite{ROCKEL20171, ROCKEL2016666} compared the wake turbulence characteristics of single and twin floating wind turbines under low- and high-turbulence inflow conditions and found that linear-surge motions led to suppressed entrainment of kinetic energy into the wake and therefore decreased wake recovery. Wei and Dabiri \cite{Wei2022phase, Wei_Dabiri_2023}, using analytical models and experiments, showed that surge platform motion can generate over $6\%$ more power under certain loading conditions than stationary turbines. In a recent study by the same group \cite{Wei_El}, at $Re = 3 \times 10^5$ with surge velocity amplitudes up to $4\%$ of the mean inflow velocity, large enhancements in wake recovery relative to the steady-flow case were observed, with wake length reductions up to $46.5\%$ and improvements in available power at $10D$ downstream of up to $15.7\%$. More recently, Messmer et al.\cite{Messmer_Hölling_Peinke_2024}, using a turbine mounted on an actuated Stewart platform, showed increases in wake recovery in excess of $20\%98$ relative to the fixed turbine case as a function of streamwise (surge), and transverse side-to-side (sway) motions.  

Using porous discs to investigate the wakes of floating wind turbines within boundary layer flows, Schliffke et al.\cite{Schliffke_2020} found that at $4.6D$ downstream, surge motion did not alter the mean wake velocity, but it did affect turbulence intensity and turbulent kinetic energy profiles. In a recent study, Schliffke et al.\cite{wes-9-519-2024} showed that harmonic platform motion produces distinct frequency signatures in far-wake spectra, whereas broadband motion leaves no easily discernible marks. Additionally, van den Broek et al.\cite{BROEK20237656} and van den Berg et al.\cite{wes-8-849-2023}, using free-vortex wake simulations, demonstrated that improvements in wake recovery due to dynamic induction control strategies can, in some cases, be mitigated by turbine motions, reducing the overall effect on recovery. Wang et al.\ \cite{WANG2024131788}, employing an actuator line model (ALM) and modal analysis, reported that high surge frequencies of around $16Hz$ reduce velocity fluctuation regions, whereas great surge amplitudes of $8m$ increase velocity-deficit fluctuations, showing that surge motion leads to faster wake recovery compared with a fixed HAWT (horizontal-axis wind turbine).  

Besides surge and pitch (i.e. fore–aft) motions, Li et al.\cite{li2022onset} found that sway motion of floating wind turbines can efficiently trigger wake meandering by exciting shear-layer instabilities. When the motion frequency is close to its optimal value, defined as the frequency that obtains an earlier wake recovery, even a small motion amplitude of only $0.01D$ can result in significant far-wake meandering and recovery, increasing the wake velocity by more than 20\% of the inflow. For the case of a single WT, they observed that the optimal sway frequency generating the earliest wake recovery was close to $St \approx 0.3$. Strouhal numbers in the range $0.2 < St < 0.6$, were capable of generating some noticeable wake recovery effect. This finding was confirmed by Messmer et al.\cite{Messmer_Hölling_Peinke_2024} and Li et al.\cite{10.1063/5.0216602} for uniform inflow at the turbine-array level, though they observed that the sensitive frequency range for the array extends down to $St = 0.125$, below the $St > 0.2$ limit observed for a single WT wake. This difference was attributed to changes in shear-layer stability when upstream and downstream wakes are superimposed, highlighting sway motion as an effective wake control strategy. Altogether, these mixed results show that a better understanding of the dynamics of unsteady flows in FOWTs is needed \cite{Fontanella_2022}, and wake interactions among multiple floating wind turbines remain to be fully studied.

In view of the current literature, these studies indicate that there is a wide range of interactions between incoming flow scales and wind turbine wakes. Despite extensive work on the effects of turbulence intensity on wind turbines \cite{li2024impacts, hodgson2025impact}, the spectral characterization of the inflow has been addressed to a much lesser extent and has only recently (around 2020) received attention  \cite{mao2018far, hodgson2025impact, biswas2024effect, gambuzza2023influence, Andersen_2024, bourhis2025impact}. The development of the wake is often assumed to be a function of turbulence intensity alone, while the spectral characteristics of the incoming flow are often ignored \cite{gambuzza2023influence, stanislawski2023effect}. Blackmore et al.\cite{blackmore2014influence} and Ghate et al.\cite{Ghate2018} have shown that inflow turbulence intensity has limited effects on wake width, whereas increasing the inflow integral length scale broadens the wake (promoting wake breakdown). 
It is therefore apparent that a better understanding of the fluid mechanics underlying the scale-dependent nature of inflow turbulence and turbine–wake interactions is still needed \cite{li2024impacts}, particularly in the case of multiple FOWTs, where changes in the spectral content of wakes remain a significant challenge. While recent studies have highlighted how atmospheric stability alters wake recovery and blockage in large wind farms \cite{Souaiby2025Atmospheric}, the role of inflow spectral scales and their coupling with FOWT platform motions has not yet been clarified.

Therefore, the present work aims to address this gap by quantifying the effect of inflow turbulent length scales on the wake dynamics of floating offshore wind turbines (FOWTs) under controlled conditions of identical turbulence intensity and mean velocity. The NREL-5MW reference turbine \cite{Jonkman2009} is selected as the turbine model because of its extensive documentation and established use as a megawatt-scale benchmark.  

Specifically, we investigate the persistence of inflow turbulent scales and their role in wake recovery for two FOWTs arranged in tandem at a spacing of $5D$, under both uniform inflow and turbulent inflow. Beyond inflow conditions alone, we examine how platform surge motions interact with these inflow scales. In particular, we also explore cases where both rotors undergo surge motion and evaluate the effect of the phase difference $\Delta \phi$ between their motions.

This study directly addresses two key research needs: (i) understanding how spectral properties of turbulence, beyond turbulence intensity, govern wake entrainment and recovery in multi-turbine configurations, and (ii) clarifying how FOWT platform motions couple with these turbulent scales to shape wake development. By disentangling these effects, the study provides new insight into the physics of unsteady wake–turbulence interactions in floating wind farms. The results are expected to enhance both the design of offshore wind farms and the development of flow-control strategies that exploit turbulent length scales and platform dynamics to increase energy capture.

\section{Numerical Methodology}
\label{sec2}

\subsection{Governing Equations and LES Turbulence Modeling}

The governing equations, which are the filtered three-dimensional incompressible Navier–Stokes equations with the Boussinesq hypothesis applied to model the sub-grid-scale stresses (SGS), are expressed as follows
\begin{equation}
\nabla \cdot \boldsymbol{u} = 0
\end{equation}
\begin{equation}
\begin{aligned}
\frac{\partial \boldsymbol{u}}{\partial t} + (\boldsymbol{u} \cdot \nabla)\boldsymbol{u} &=  \frac{\nabla p}{\rho} + \nabla \cdot \Big[ (\nu + \nu_{\mathrm{SGS}}) \left( \nabla \boldsymbol{u} + \nabla \boldsymbol{u}^{T} \right) \Big] + \frac{\boldsymbol{F}}{\rho}
\end{aligned}
\label{eq:N-S}
\end{equation}
where \(\boldsymbol{u}\), \(p\), $\boldsymbol{F}$,  \(\rho\), \(\nu\), and \(\nu_{\mathrm{SGS}}\) denote the (filtered) velocity vector, (filtered modified) pressure, (filtered) body force vector, fluid density, kinematic viscosity, and SGS viscosity. Note that the flow is treated as Newtonian, that is, the kinematic viscosity \(\nu\) is constant.  

In the present work, the Wall-Adapting Local Eddy-Viscosity (WALE) model \cite{nicoud1999subgrid} is employed as the subgrid-scale closure for LES, owing to its robustness in capturing near-wall turbulence and transitional features on structured meshes. Turbulence closures approximate the deviatoric part of the SGS stress tensor through the Boussinesq hypothesis using the sub-grid-scale viscosity \(\nu_{\mathrm{SGS}}\), which is given by:

\begin{equation}
\nu_{\mathrm{SGS}} = \left( C_w \Delta \right)^2 \frac{\left( \mathcal{S}^d : \mathcal{S}^d \right)^{3/2}}{\left( \mathbf{S} : \mathbf{S} \right)^{5/2} + \left( \mathcal{S}^d : \mathcal{S}^d \right)^{5/4}}
\end{equation}

\begin{equation}
\mathcal{S}^d = \frac{1}{2} \left[ \left( \nabla {\boldsymbol{u}} \right)^2 + \left( \nabla {\boldsymbol{u}}^T \right)^2 \right] - \frac{1}{3} \left( \nabla \cdot {\boldsymbol{u}} \right)^2 \mathbf{I}
\end{equation}
here \(\Delta\) is the filtering length, \(\mathbf{I}\) is the identity tensor, \(\mathbf{S} = \tfrac{1}{2} \left( \nabla {\mathbf{u}} + \nabla {\mathbf{u}}^T \right)\) denotes the (filtered) strain-rate tensor, and \(\mathcal{S}^d\) represents the traceless symmetric part of the squared (filtered) velocity-gradient tensor.

The filter width in the current work is defined as \(\Delta = \sqrt{\Delta x \, \Delta y \, \Delta z}\), with \(\Delta x, \Delta y, \Delta z\) denoting the local grid spacings along the coordinate directions. The constants adopted in the WALE formulation are \(C_e = 1.048\), \(C_k = 0.094\), and \(C_w = 0.325\), following established values from the literature \cite{nicoud1999subgrid, shukla2019flow}.  
By accounting for both strain- and rotation-related effects, the WALE model provides enhanced accuracy in representing wake dynamics within the LES framework \cite{nabhani5414004high}.

\subsection{Surge-Aware Actuator Line Method}

\begin{figure*}[htbt]
    \centering
    \includegraphics[width=0.95\textwidth]{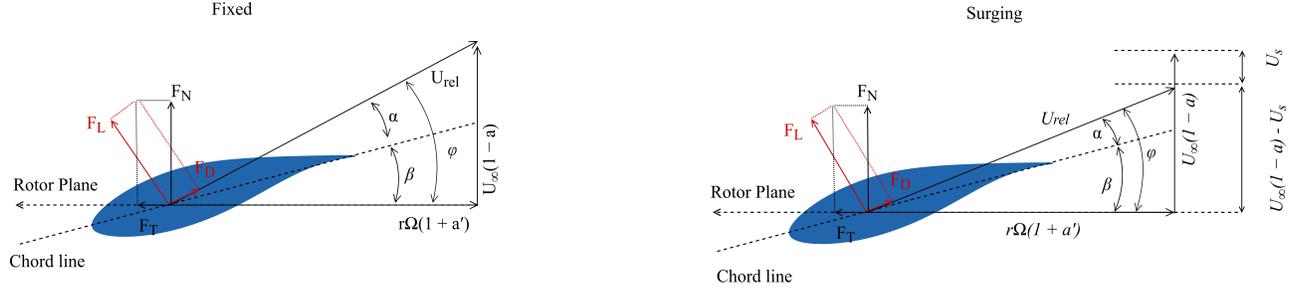}
    \caption{Velocity triangle and force decomposition on a blade element for fixed (left) and surging (right) turbine conditions. Surge motion introduces an additional axial velocity component $U_s$, which alters the apparent inflow velocity $U_{\text{rel}}$, the inflow angle $\phi$, and consequently the angle of attack $\alpha$. }
    \label{fig:U-diagram}
\end{figure*}

The aerodynamic response of floating offshore wind turbines (FOWTs) undergoing surge motion is captured here using a surge-aware extension of the Actuator Line Method (ALM). This implementation builds on the open-source \texttt{turbinesFoam} framework \cite{bachant2016turbinesfoam}, which is modified and extended for the present study and referred to as \texttt{SurgeMotionFoam}. 
In this framework, the turbine blades are represented as lines of rotating actuator points that exert body forces on the flow. These actuator points also translate according to the FOWT’s motion. The body forces exerted by the actuator points are computed from the local apparent flow velocity sampled at each actuator point and the local sectional properties of the turbine blades, as described below.
The distinctive feature of the surge-aware formulation is the explicit inclusion of the platform’s surge velocity $U_s$, which affects the apparent inflow seen by the rotating blades of the wind turbines. As shown in Fig.~\ref{fig:U-diagram}, surge motion affects the axial component of the relative velocity ${U}_{\text{rel}}$, thereby influencing the inflow angle $\phi$ and the angle of attack $\alpha$ of a sectional element. The resulting relative velocity magnitude is expressed as
\begin{equation}
    U_{\text{rel}} = \sqrt{ \left[ (1 - a) U_\infty - U_s \right]^2 + \left[ \Omega r (1 + a') \right]^2 }
    \label{eq:urel}
\end{equation}
where $a$ and $a'$ denote the axial and tangential induction factors, respectively, and $r$ is the local radial distance to the rotor center.

After sampling the relative velocities seen by the actuator points, the sectional aerodynamic forces can be evaluated as
\begin{equation}
    {\boldsymbol{f}}_{2D}(\boldsymbol{r}) = (L, D) =  \tfrac{1}{2} \rho c U_{\text{rel}}^2 
    \left(C_l(\alpha)\, \hat{\mathbf{e}}_L + C_d(\alpha)\, \hat{\mathbf{e}}_D\right)
    \label{eq:f2d}
\end{equation}
where $\boldsymbol{r}$ is the position vector of the actuator point, $c$ is the local chord length of the blade, $C_l(\alpha)$ and $C_d(\alpha)$ are the lift and drag coefficients, and $\hat{\mathbf{e}}_L$, $\hat{\mathbf{e}}_D$ are the unit vectors in the lift and drag directions, respectively. These forces are then projected into normal and tangential components with respect to the rotor plane, providing thrust and torque contributions at each actuator element.

The body force field $\boldsymbol{F}$ appearing in Equation~\eqref{eq:N-S}, is obtained by mapping the aerodynamic forces ${\boldsymbol{f}}_{2D}(\boldsymbol{r})$ onto the flow field via a smoothing kernel $\eta_\epsilon$:
\begin{subequations}
\label{eq:body_force}
\begin{equation}
\begin{aligned}
\boldsymbol{F} &= \sum_{i=1}^{B} \int_{0}^{R} f_{\text{tip}}(r_i)\, \boldsymbol{f}_{2D}(r_i)\, \eta_\epsilon\!\left( \left\| \boldsymbol{x} - \left( r_i \hat{\mathbf{e}}_i + \boldsymbol{p}_R \right) \right\| \right) \, \mathrm{d} r_i
\end{aligned}
\end{equation}
\begin{equation}
\eta_\epsilon(d)
= \frac{1}{\epsilon^{3}\pi^{3/2}}
\exp\!\left[
- \left(\frac{d}{\epsilon}\right)^{2}
\right]
\end{equation}
\end{subequations}

where $\boldsymbol{x}$ is the position vector, $B$ is the number of blades, $\boldsymbol{p}_R$ is the rotor reference position, and $f_{\text{tip}}(r)$ denotes the Glauert-type tip correction factor with its expression being
\begin{equation}
    f_{\text{tip}}(r) = \frac{2}{\pi} \arccos \left[ 
    \exp\!\left( \frac{-B(R - r)}{2r \sin\phi} \right) \right]
    \label{eq:ftip}
\end{equation}
The tip correction factor $f_{\text{tip}}(r)$ is applied to ensure that the aerodynamic loading smoothly diminishes to zero near the blade tip \cite{TroldborgALM2009, jha2014guidelines}.
In the present study, each blade is discretized into 40 actuator points distributed along the span. Following the established ALM practices \cite{SARLAK2015386}, the Gaussian smoothing width is set to $\epsilon = 2\Delta$, ensuring smooth and consistent projection of aerodynamic loads \cite{TroldborgALM2009}. Note that the hub is modeled as an additional actuator element with a drag coefficient $C_d = 0.3$ \cite{NADERI20171346}, the hub radius being $1.5$~m \cite{Jonkman2009}. 
It is worth noting that, for the sake of simplicity, dynamic stall and aero-elastic effects are not included in this implementation. A comprehensive theoretical discussion of surge-aware actuator modeling can be found in Li et al.~\cite{li2024wake, LI2025122062}.

\subsection{Surge Settings}

\begin{figure*}[htbp]
    \centering
    \includegraphics[width=0.95\textwidth]{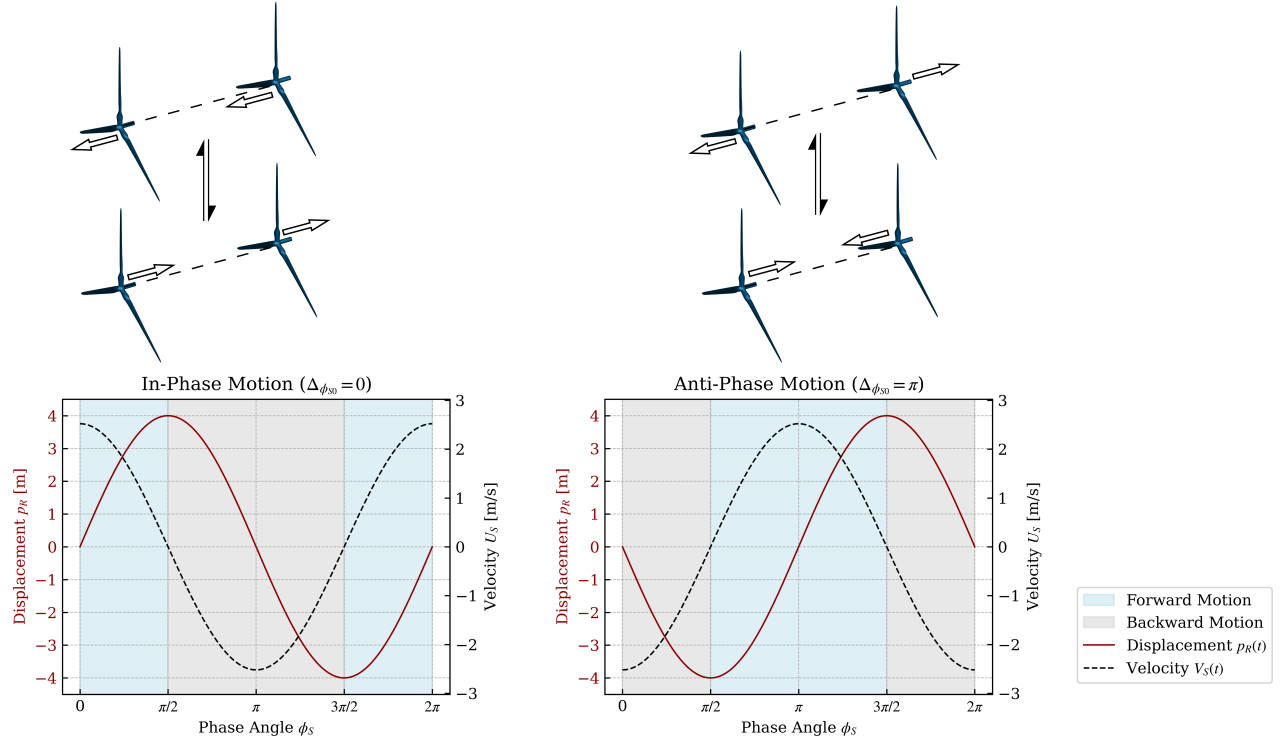}
    \caption{Illustration of in-phase ($\Delta\phi_{S0} = 0$) and anti-phase ($\Delta\phi_{S0} = \pi$) surge motions of tandem floating wind turbines. 
    The top panels depict the relative surge alignment of the upstream and downstream turbines. 
    The bottom plots show the corresponding displacement $p_R(t)$ and surge velocity $U_S(t)$ as functions of the normalized surge phase angle $\phi_S$.} 
    \label{fig:phase_motion}
\end{figure*}

Floating offshore wind turbines are prescribed with sinusoidal surging motions with amplitudes $A_{\text{sur}}^{\text{up}}$ and $A_{\text{sur}}^{\text{dn}}$ and angular frequencies $\omega_{\text{sur}}^{\text{up}}$ and $\omega_{\text{sur}}^{\text{dn}}$ for the upstream and downstream turbines, respectively. Unless stated otherwise, the surge amplitude is set to $4$~m and the frequency to $0.63$~rad/s. These sets of values are consistent with realistic platform responses under site-specific wave forcing \cite{li2024wake, LI2025122062, hansen2012impact}.

The instantaneous position of the upstream ($\boldsymbol{p}_R^{\text{up}}$) and downstream ($\boldsymbol{p}_R^{\text{dn}}$) rotors is expressed as:

\begin{align}
\boldsymbol{p}_R^{\text{up}}(t) &= \boldsymbol{p}_R^{\text{up},0} + A_{\text{sur}}^{\text{up}} \sin\!\left(\omega_{\text{sur}}^{\text{up}} t + \phi_0^{\text{up}}\right) \hat{\mathbf{e}}_x \\
\boldsymbol{p}_R^{\text{dn}}(t) &= \boldsymbol{p}_R^{\text{dn},0} + A_{\text{sur}}^{\text{dn}} \sin\!\left(\omega_{\text{sur}}^{\text{dn}} t + \phi_0^{\text{dn}}\right) \hat{\mathbf{e}}_x
\end{align}
where $\hat{\mathbf{e}}_x$ is the streamwise unit vector and $\phi_0$ the initial phase offset (see Figure~\ref{fig:phase_motion}). The relative phase difference between upstream and downstream surge motions is defined as
\begin{equation}
\Delta \phi = \phi_0^{\text{dn}} - \phi_0^{\text{up}}
\end{equation}

The corresponding surge velocities $U_S$ take the following form
\begin{align}
U_S^{\text{up}}(t) &= A_{\text{sur}}^{\text{up}} \, \omega_{\text{sur}}^{\text{up}} \cos\phi_S^{\text{up}} \\
U_S^{\text{dn}}(t) &= A_{\text{sur}}^{\text{dn}} \, \omega_{\text{sur}}^{\text{dn}} \cos\phi_S^{\text{dn}}
\end{align}

All simulations are carried out at the rated operating condition of the upstream turbine, corresponding to a freestream velocity $U_\infty = 11.4$~m/s and a rotational speed $\Omega = 1.27$~rad/s, which leads to a tip speed ratio TSR $= 7$. Identical operating conditions are imposed on the downstream rotor. The surge frequency is chosen such that $\omega_{\text{sur}} / \Omega \approx 2\pi/3$, ensuring consistent phase resolution across complete rotor revolutions. The remaining degrees of freedom are assumed to be fixed. 

Cycle-averaged performance metrics are also computed, including the power and thrust coefficients:
\begin{align}
C_P &= \frac{P^*}{\tfrac{1}{2} \rho \, {U_\infty^3} A} \\
C_T &= \frac{T}{\tfrac{1}{2} \rho \, {U_\infty^2} A}
\end{align}
where $P^*$ represents the power contributions from torque and surge-induced motion, $T$ is the rotor thrust, and $A$ characterizes the turbine swept area.  
To ensure statistical convergence of both phase- and cycle-averaged quantities, after achieving convergence, over 60 complete rotor revolutions were used for each simulated case to post-process the information presented in this manuscript.

\subsection{Inflow Turbulence Modeling}

To reproduce realistic atmospheric inflow conditions, unsteady turbulence was imposed at the inlet using the divergence-free synthetic eddy method (DFSEM) \cite{poletto2011divergence, poletto2013new}. This method generates velocity fluctuations that are spatially and temporally correlated while enforcing incompressibility, thereby ensuring mass conservation.
Offshore atmospheric boundary layers generally exhibit weaker turbulence than over land due to reduced surface roughness. To reflect offshore conditions, we prescribed an inlet turbulence intensity (TI) of $5.3\%$ for all turbulent cases, consistent with field observations and prior modeling \cite{hansen2012impact}. 
The inflow turbulence intensity is defined as:

\begin{equation}
TI = \frac{u'_{RMS}}{\langle {U}\rangle} = \frac{\sqrt{\frac{1}{3}(\langle{u'u'}\rangle + \langle{v'v'}\rangle + \langle{w'w'}\rangle)}}{ {\sqrt{U_x^2+U_y^2+U_z^2}}}
\label{turb_intensity}
\end{equation}

where $u'$, $v'$, and $w'$ denote the velocity fluctuations in the streamwise, vertical, and lateral directions, respectively, $u'_{RMS}$ denotes the Root-Mean-Square (RMS) of the streamwise velocity fluctuations, and $\langle {U}\rangle$ is the mean hub-height velocity. The $\langle ... \rangle$ represents the operator for time-averaging. Owing to the streamwise evolution of the synthetic field and scale-dependent production/dissipation, the realized TI at hub height is observed to vary (see section \ref{TI-initial-result}) with the imposed integral length scale $L_u$.
To investigate the role of the turbulence structure, five distinct integral streamwise length scales $L_u$ were prescribed at the inlet. 
The streamwise integral length scale $L_u$ characterizes the size of the dominant eddies and is obtained from the autocorrelation function of the streamwise velocity fluctuations $R_{uu}(\tau)$ as:
\begin{equation}
L_u = \int_0^\infty R_{uu}(\tau)\,d\tau
\label{length_scale}
\end{equation}
where
\begin{equation}
R_{uu}(\tau) = \frac{\langle {u'(x_0,t)\,u'(x_0 + \tau,t)\rangle}}{\langle{{u'(x_0,t)}^2}\rangle}
\label{two_point_correlation}
\end{equation}
The parameter $\tau$ defines the streamwise separation distance between consecutive mesh cells. The term ${u'(x_0,t)}$ represents the instantaneous velocity fluctuation in the streamwise direction measured at the position $x_0$ and at a given time ($t$). The term $\langle {{u'(x_0,t)}^2} \rangle$ characterizes the variance (the square of the standard deviation) of the streamwise velocity fluctuations. 
In practice, the divergence-free synthetic eddy method (DFSEM) enables $L_u$ to be directly specified, allowing systematic variation of the inflow coherence and turbulence scale across all cases.
The following non-dimensional length scales relative to the rotor radius $R$ were used:
\[
L_u/R = \{0.25,\ 0.50,\ 0.75,\ 1.00,\ 1.25\}.
\]
\
This range spans fine-scale turbulence ($L_u/R=0.25$) to large-scale structures comparable to those observed in stable offshore boundary layers ($L_u/R=1.25$). Varying $L_u$ at fixed turbulence intensity at the computational domain inlet pretends to isolate the influence of eddy size on wake dynamics and turbine performance.
In all cases, turbulent fluctuations were superimposed on a steady inflow profile with hub-height velocity $U_\infty = 11.4~\mathrm{m/s}$. The perturbations were introduced at the inlet and advected downstream with the mean flow. The flow characteristics were investigated by examining the TI, energy spectra, and correlation lengths in a given downstream location. 

\section{Simulation setup and Grid Evaluation}

\subsection{Simulation framework}

All simulations were conducted using \texttt{OpenFOAM} version \texttt{v2312} \cite{OpenFOAM-v2312}, employing the \texttt{pimpleFoam} solver for transient, incompressible, turbulent flow. The air density was set to $\rho = 1.225~\mathrm{kg/m^3}$, and the kinematic viscosity to $\nu = 1.5 \times 10^{-5}~\mathrm{m^2/s}$. A Large-eddy simulation (LES) approach was applied with the WALE subgrid-scale model as previously described.
Second-order accuracy was used for both spatial and temporal discretizations. A central differencing scheme (Gauss linear) was used for gradient and divergence terms, while the time integration was handled using the Crank--Nicolson method with a blending factor of $0.9$. All solver residuals were set to a convergence criterion of $10^{-6}$, and the maximum Courant–Friedrichs–Lewy's number was limited to $0.7$ throughout the simulations to ensure numerical stability.

The simulations were executed on the high-performance computing (HPC) facility \textbf{MareNostrum 5 GPP} at the Barcelona Supercomputing Center (BSC) \cite{BSC_MareNostrum5}. Each computing node comprises two Intel Xeon Platinum 8480+ processors (56 cores per socket), totaling 112 cores per node, clocked at 2.0~GHz. Each case was run for a total physical time of 600 seconds, corresponding to approximately 120 rotor revolutions. Each revolution was discretized into 360 time steps, resulting in high temporal resolution. The simulations were parallelized over 10 nodes (1,120 cores total) and completed in under 24 hours per case. 

\subsection{Mesh generation and quality assessment}

\begin{figure}
    \centering
    \begin{subfigure}[b]{0.49\textwidth}
        \centering
        \includegraphics[width=\linewidth]{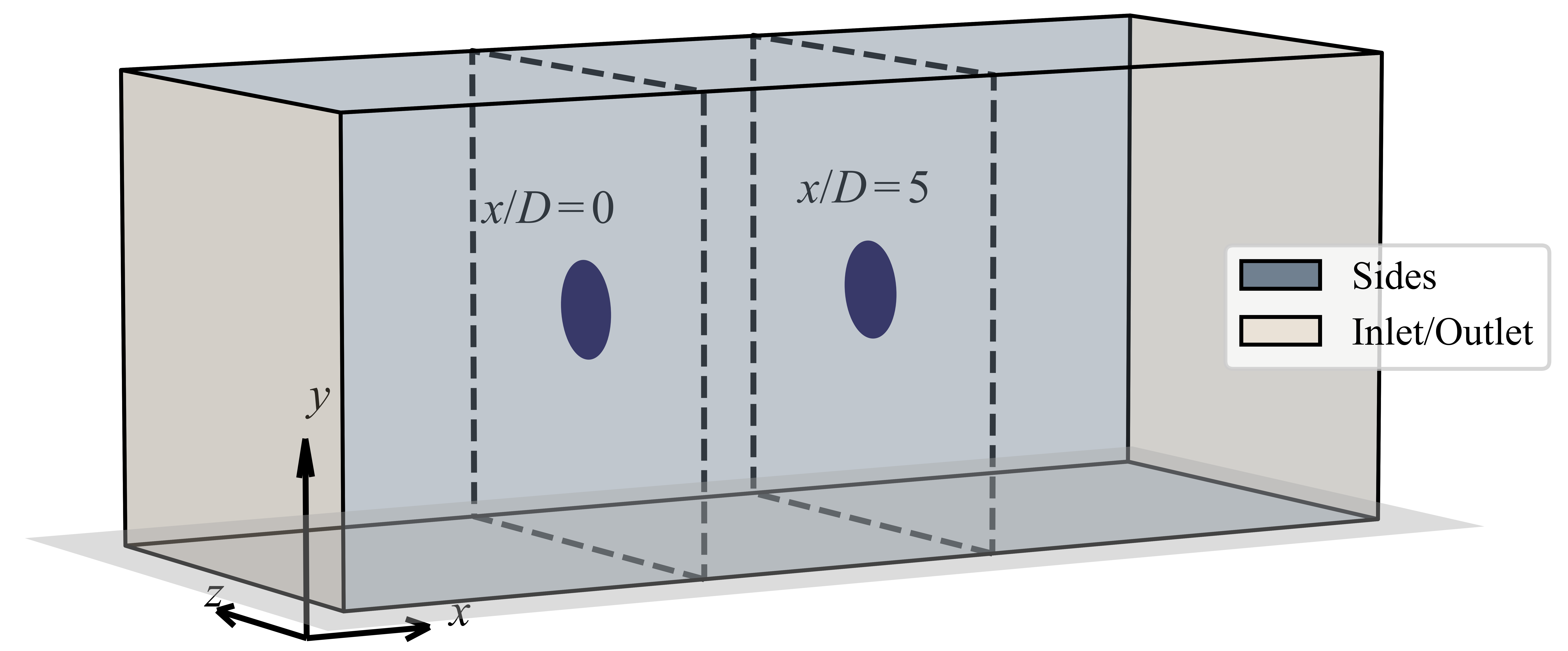}
        \caption{Computational domain}
        \label{fig:cmp-domain}
    \end{subfigure}
        \centering
    \begin{subfigure}[b]{0.49\textwidth}
        \centering
        \includegraphics[width=\linewidth]{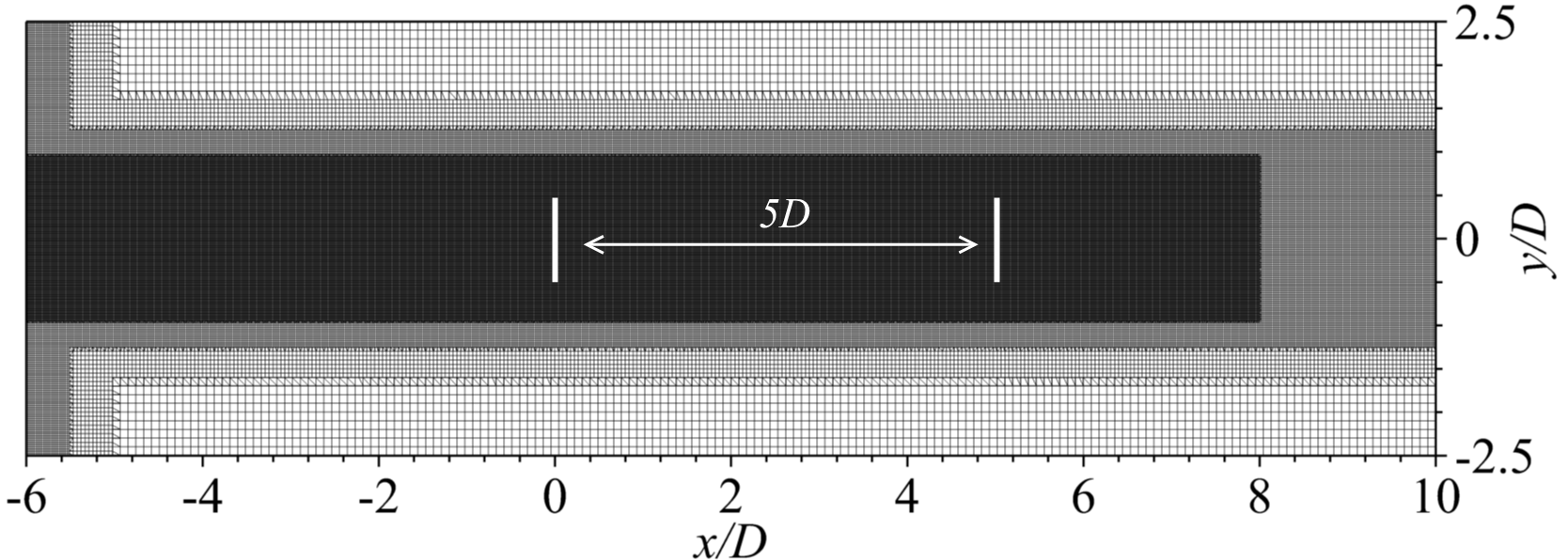}
        \caption{Computational mesh}
        \label{fig:cmp-mesh}
    \end{subfigure}
    \caption{(a) Overall computational domain with the location of the upstream and downstream turbines. (b) Computational mesh shown in an $x$–$y$ plane. Three levels of grid refinement are applied in the region of interest. The background mesh is generated using \texttt{blockMesh}, while local refinement around the rotors is performed using \texttt{snappyHexMesh}.}
    \label{fig:mesh}
\end{figure}

\begin{figure*}[ht!]
    \centering
    \includegraphics[width=\linewidth]{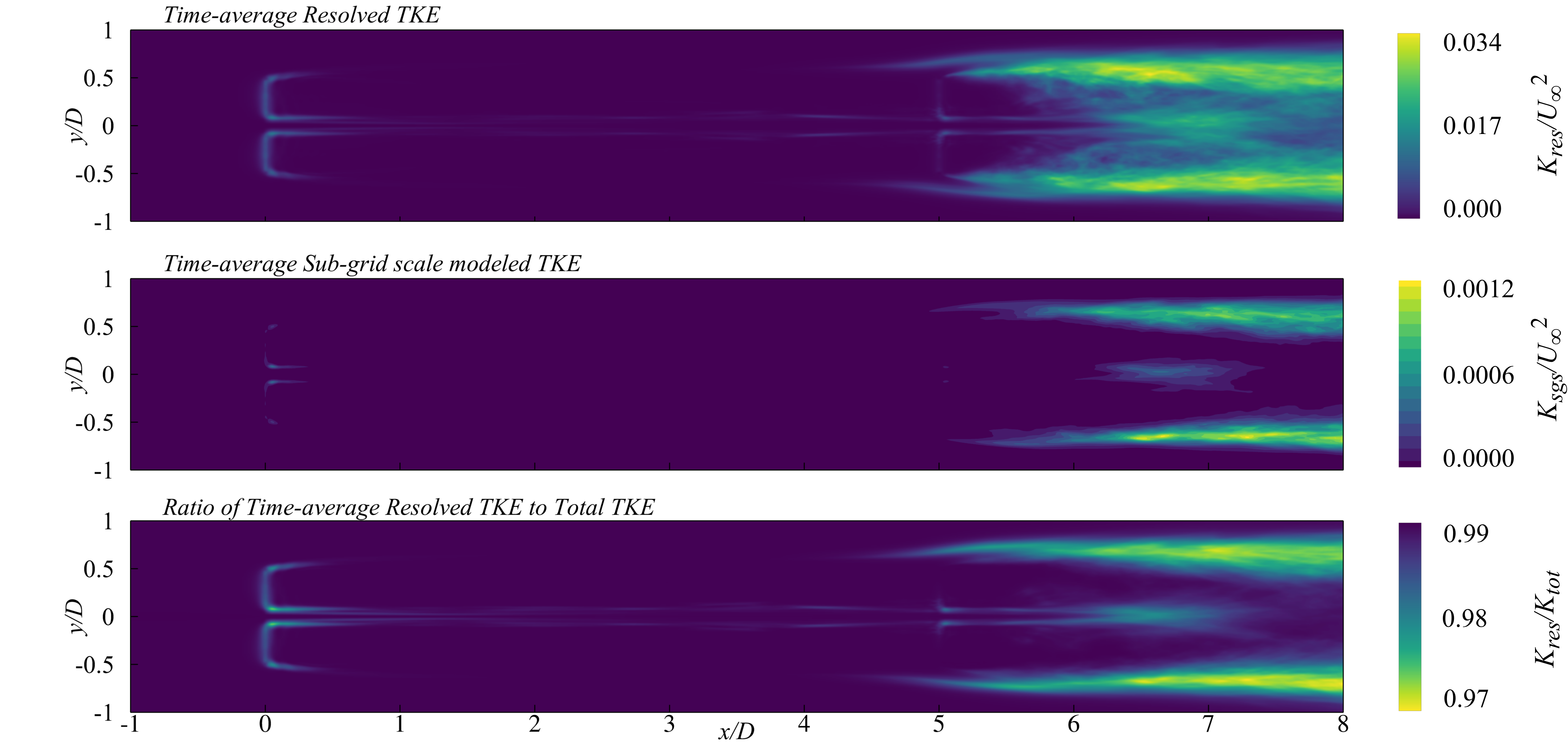}
    \caption{Assessment of LES resolution quality: (top) resolved turbulent kinetic energy $K_\text{res}$; (middle) SGS-modeled TKE $K_\text{SGS}$; (bottom) ratio $K_\text{res}/(K_\text{res} + K_\text{SGS})$ demonstrating the dominance of resolved turbulence.}
    \label{fig:TKE-ratio}
\end{figure*}

\begin{figure*}
    \centering
    \begin{subfigure}[b]{0.48\textwidth}
        \centering
        \includegraphics[width=\linewidth]{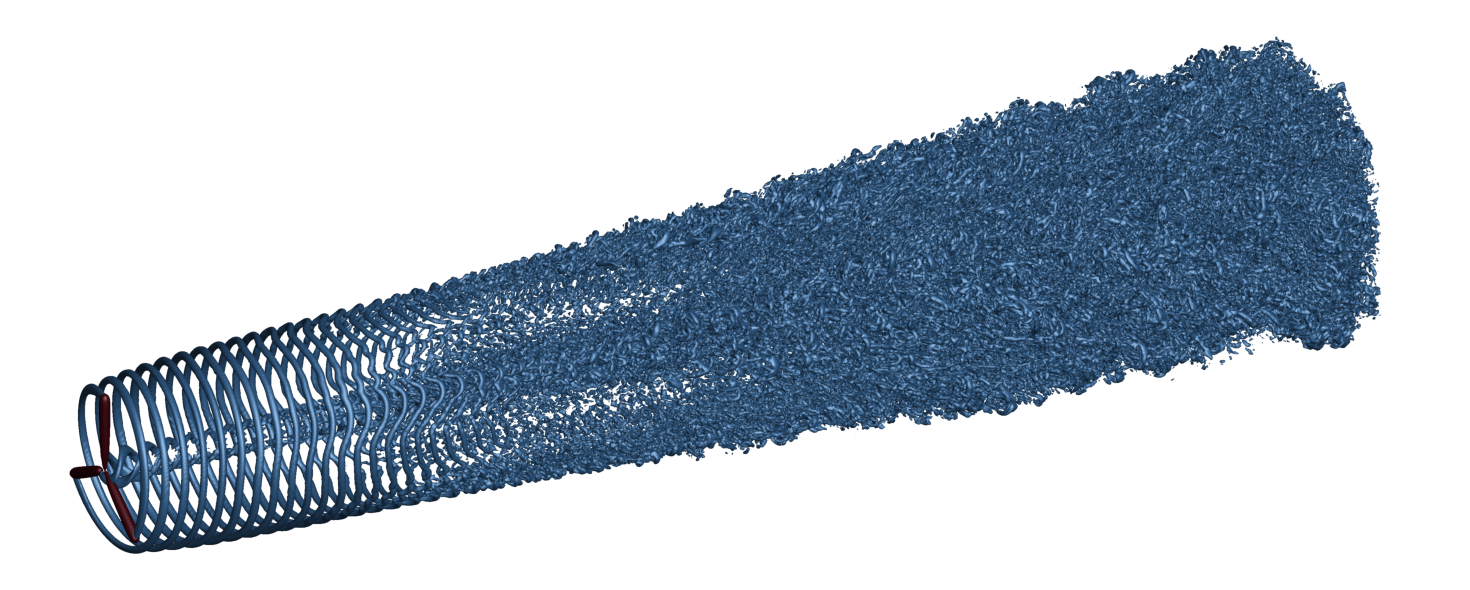}
        \caption{Q iso-surface (fixed case)}
        \label{fig:q_iso_fix}
    \end{subfigure}
    \begin{subfigure}[b]{0.48\textwidth}
        \centering
        \includegraphics[width=\linewidth]{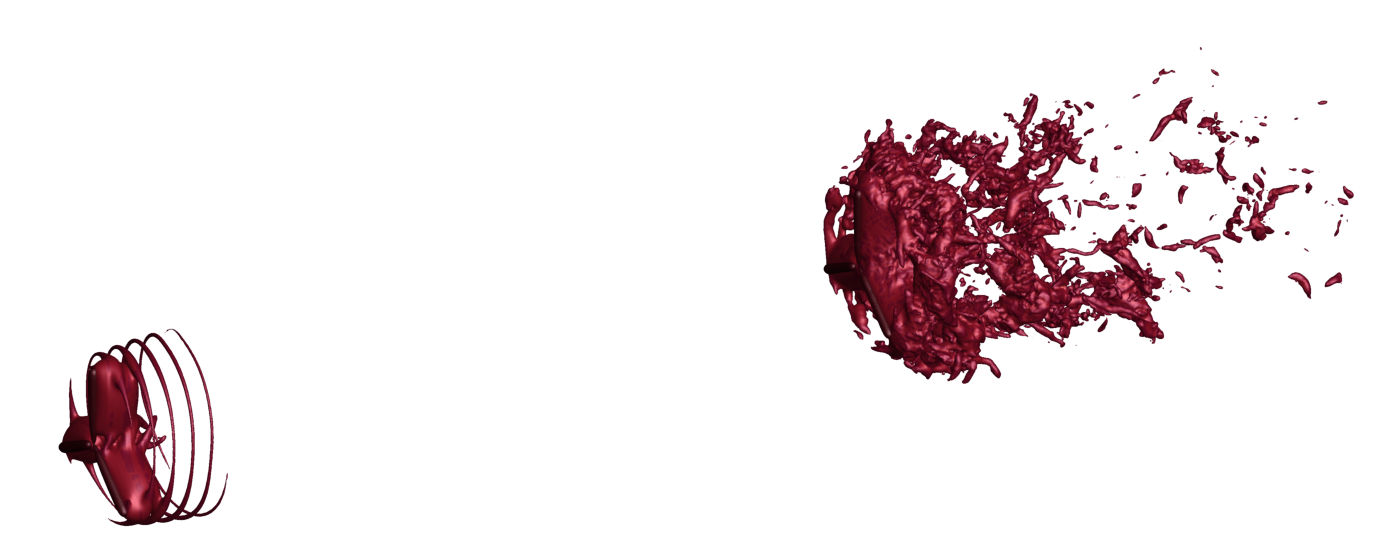}
        \caption{p iso-surface (fixed case)}
        \label{fig:p_iso_fix}
    \end{subfigure}
    \begin{subfigure}[b]{0.48\textwidth}
        \centering
        \includegraphics[width=\linewidth]{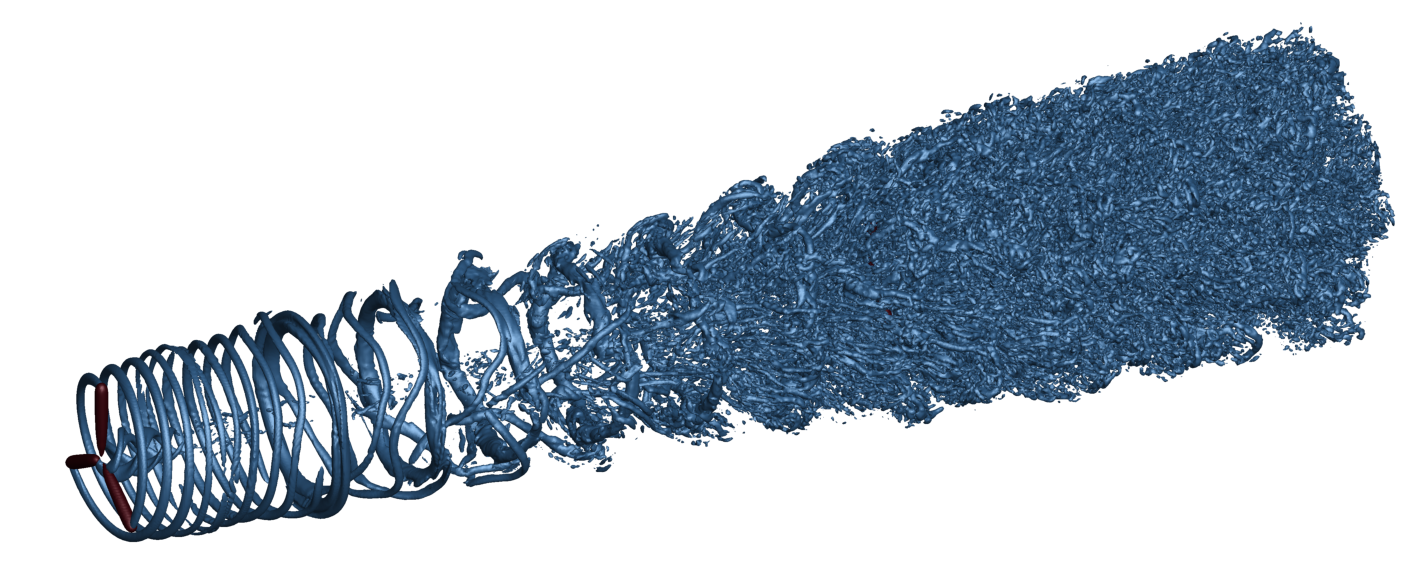}
        \caption{Q iso-surface (surge case)}
        \label{fig:q_iso_surg}
    \end{subfigure}
    \begin{subfigure}[b]{0.48\textwidth}
        \centering
        \includegraphics[width=\linewidth]{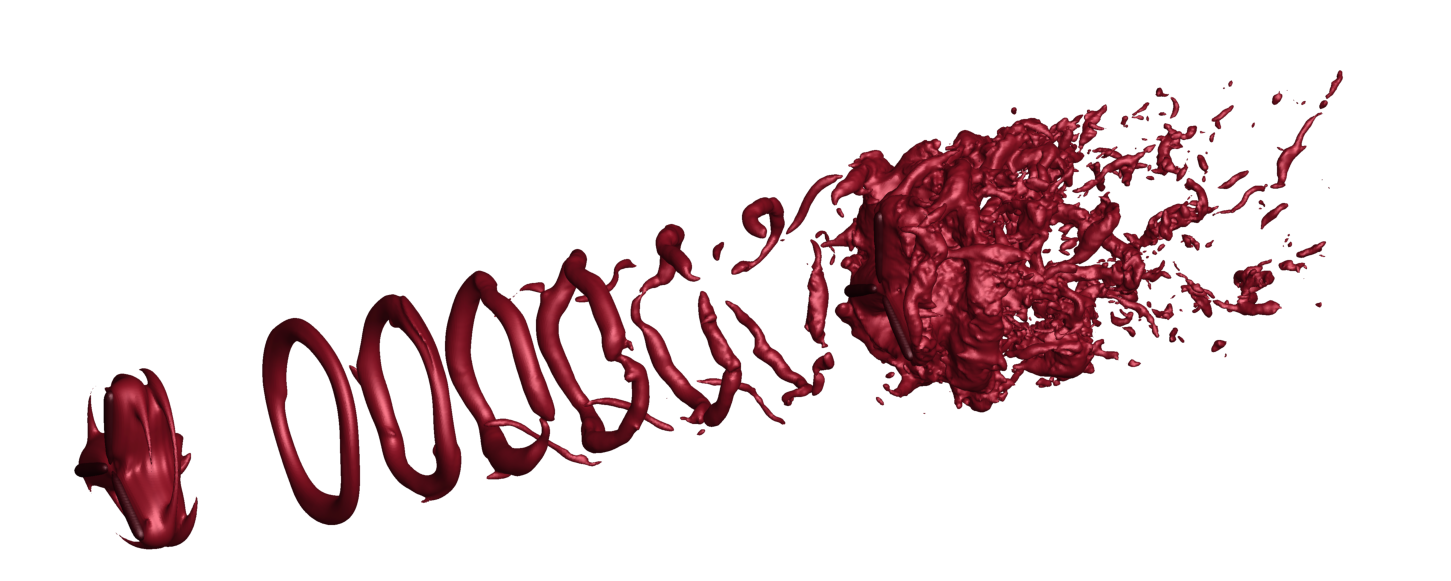}
        \caption{p iso-surface (surge case)}
        \label{fig:p_iso_surg}
    \end{subfigure}
    \caption{Iso-surface visualizations of the instantaneous kinematic pressure ($p=\Delta p/\rho = -8$~m$^2$/s$^2$) and Q criterion ($Q=0.25$~s$^{-2}$) fields for both fixed and surge cases.  Sub-figures (a) and (b) correspond to the fixed configuration, while (c) and (d) illustrate the surge configuration, the upstream turbine prescribes surge motion while the downstream one remains fixed.}
    \label{fig:iso_surfaces}
\end{figure*}

The computational domain employed for all simulation cases is depicted in Figure~\ref{fig:mesh}. A consistent mesh topology, comprising roughly 94 million cells, was maintained across all configurations to ensure comparability of results. The mesh was constructed in two stages: a background mesh using \texttt{blockMesh}, followed by local refinement using three successive refinement levels with \texttt{snappyHexMesh}. The highest level of refinement in the vicinity of the rotor blades achieved a grid spacing of approximately $\Delta x \approx 6.3 \times 10^{-3}$~D, sufficient to capture fine-scale vortex structures and atmospheric boundary layer behavior \cite{yen2024near}. 
The domain itself spanned $16D$ in the streamwise ($x$) direction, and $5D$ in both the vertical ($z$) and lateral ($y$) directions, where $D$ represents the rotor diameter. The upstream turbine ($T^{up}$) was placed at $x = 0.0D$, $6D$ downstream of the computational domain inlet, allowing adequate distance from the inlet boundary to establish a stabilized inlet velocity profile. A second rotor ($T^{down}$) was located at $x = 5.0D$, yielding an inter-turbine spacing of $5D$ center-to-center. This positioning ensures that $T^{down}$ experiences a developed wake field influenced by the upstream rotor, while avoiding the immediate near-wake dominated by unstable and transient vortex shedding.
The boundary conditions were defined to ensure both physical realism and numerical stability throughout the domain. At the inlet, a fixed-value (Dirichlet) condition was imposed on the velocity field to prescribe the inflow profile, while a zero-gradient (Neumann) condition was applied for pressure. At the outlet, the pressure was fixed to a reference value of $p = 0$ (Dirichlet), and a zero-gradient condition was used for the velocity field to allow for fully developed outflow. All sides were treated with no-slip boundary conditions for velocity, enforcing zero relative motion at the wall. Correspondingly, pressure was assigned a zero-gradient condition at these walls. 


To validate the adequacy of the mesh resolution in this study, contours of TKE quantities are shown in Figure~\ref{fig:TKE-ratio}. They are evaluated on the horizontal wake plane ($z = 0$) in the $(x, y)$ coordinates for the case with both turbines operating at a fixed tip speed ratio of TSR $= (\Omega \times R)/U_\infty = 7$ (where $\Omega$ is the angular velocity, and $R$ is the rotor radius). The platform motion was disabled in this analysis to isolate wake development and turbulence resolution under quasi-steady operating conditions.

The top panel shows the resolved turbulent kinetic energy ($k_\mathrm{res}$), computed as: 

\begin{equation}
    k_\mathrm{res} = \frac{1}{2} \left( \langle u'^2 \rangle + \langle v'^2 \rangle + \langle w'^2 \rangle \right),
\end{equation}

As expected, high values of $k_\mathrm{res}$ occur immediately downstream of the rotor tips and along the shear layers of the wake, with enhanced intensity in regions where coherent tip vortices begin to destabilize, indicating the onset of wake breakdown and turbulent mixing.

The middle panel presents the modeled subgrid-scale TKE ($k_\mathrm{SGS}$), defined by:

\begin{equation}
    k_\mathrm{SGS} = \frac{\nu_\mathrm{SGS}}{C_k \Delta},
\end{equation}

where $\nu_\mathrm{SGS}$ represents the eddy viscosity from the WALE model, $\Delta$ characterizes the local grid filter width based on cell volume, and $C_k = 0.094$ is a standard model constant. The SGS contribution is found to be small and localized, consistent with expectations for a well-resolved LES mesh.

The bottom panel shows the ratio of resolved to total TKE:

\begin{equation}
    k_\mathrm{ratio} = \frac{k_\mathrm{res}}{k_\mathrm{res} + k_\mathrm{SGS}},
\end{equation}

This metric serves as a quantitative indicator of LES fidelity. Across the wake field, the $k_\mathrm{ratio}$ exceeds 0.97 in the near and mid-wake regions, far surpassing the commonly accepted threshold of 0.80 for high-quality LES resolution, as suggested by Pope \cite{StephenBPope_2001, pope2004ten} and later by Tousi et al. \cite{TOUSI2022107679}. This confirms that the current mesh is sufficiently refined to resolve the spatial-temporal development of fine-scale flow structures, including the evolution of the tip vortex and rotor–wake interaction.

In conclusion, the TKE-based analysis confirms that the adopted mesh supports a reliable LES simulation. The dominance of resolved turbulence and the limited SGS contribution validate the quality of the numerical setup and the suitability of the WALE model for capturing the unsteady wake dynamics behind the floating wind turbine configuration under steady operational conditions.

Figure~\ref{fig:iso_surfaces} presents iso-surfaces of the Q-criterion and pressure fields, providing insight into the vortical structures and wake dynamics for both the static case (panels (a) and (b)) and the in-phase surge case (panels (c) and (d), sampled at t=2$\pi$, see Figure~\ref{fig:phase_motion}). Panels (a) and (c) show Q-criterion iso-surfaces, visualized with a threshold of $Q = 0.25~\mathrm{s}^{-2}$, while panels (b) and (d) display iso-surfaces of instantaneous kinematic pressure corresponding to the same cases. These visualizations capture coherent vortical structures emanating from the turbine blades and reveal key differences in wake evolution due to upstream platform motion.
The iso-surfaces in Figure~\ref{fig:iso_surfaces}(a,b) reveal classical helical tip vortices shed from each rotor, with clearly organized structures propagating downstream. The tip vortices remain relatively symmetric and coherent across several rotor diameters, and the interaction zone between the upstream and downstream wakes is well-structured, exhibiting mild undulation as the shear layers evolve.
By contrast, Figure~\ref{fig:iso_surfaces}(b) illustrates the wake dynamics when the upstream turbine undergoes prescribed surge motion while the downstream turbine remains fixed. The surge motion introduces unsteady loading on the upstream rotor, and these variations are directly reflected in the wake. As a result, the vortical structures of the upstream rotor become visibly distorted and exhibit greater curvature and stretching. Moreover, these perturbations also enhance wake meandering and trigger earlier destabilization of the tip vortices. Consequently, although the downstream turbine is held stationary, it experiences a markedly more unsteady inflow characterized by fluctuations in vortex strength and orientation.
The corresponding pressure iso-surfaces in Figures~\ref{fig:iso_surfaces}(b) and (d) further highlight the influence of surge motion. In the fixed case (b), low-pressure cores trace the coherent vortex filaments but are constrained in the near wake region, consistent with the well-ordered Q-criterion structures. In the surge-influenced case (d), the low-pressure cores are extended to regions that are further downstream compared to the fixed case. However, the shapes of these ring-like low-pressure zones are found to be highly distorted as traveling downstream, indicating the event of vortex breakdown. This aligns with the observed increase in wake unsteadiness and energy dispersion 

Overall, the comparison between the fixed and surging configurations demonstrates the sensitivity of wake topology to platform-induced motion. Even moderate surge amplitudes can induce substantial changes in vortex dynamics, with implications of influencing the wake recovery, turbine loading, and downstream performance in floating wind farm scenarios.

A direct quantitative comparison of power coefficients for large-scale wind turbines remains challenging due to the limited availability of experimental data at full scale and the wide range of numerical approaches adopted in the literature. Existing studies employ different modeling fidelities, including blade element momentum (BEM), actuator disk, actuator line, and fully resolved blade simulations, often combined with different turbulence models and inflow conditions. As a result, reported values of the power coefficient span a relatively broad range, as summarized in Table~\ref{tab:NREL5MW} in the Appendix. Despite these differences, the present results show good agreement with recent high-fidelity numerical studies. In particular, comparison with the work of Li et al.~\cite{li2024wake} (for a fixed platform configuration and uniform inflow) reveals only minor discrepancies, with deviations of approximately $0.38\%$ in the predicted power coefficient and $1.22\%$ in the thrust coefficient. These small differences fall well within the variability reported across numerical methodologies and support the robustness of the present actuator-line-based LES framework. The close agreement further indicates that the selected grid resolution, inflow characterization, and turbulence modeling approach are sufficient to capture the dominant aerodynamic response of large-scale wind turbines under the considered operating conditions.

\section{Results and discussion}

\subsection{Inflow Turbulence Characteristics Without any Turbine in the Computational Domain}
\label{TI-initial-result}

\begin{figure*}
    \centering
    \begin{subfigure}[b]{\textwidth}
        \centering
        \includegraphics[width=\linewidth]{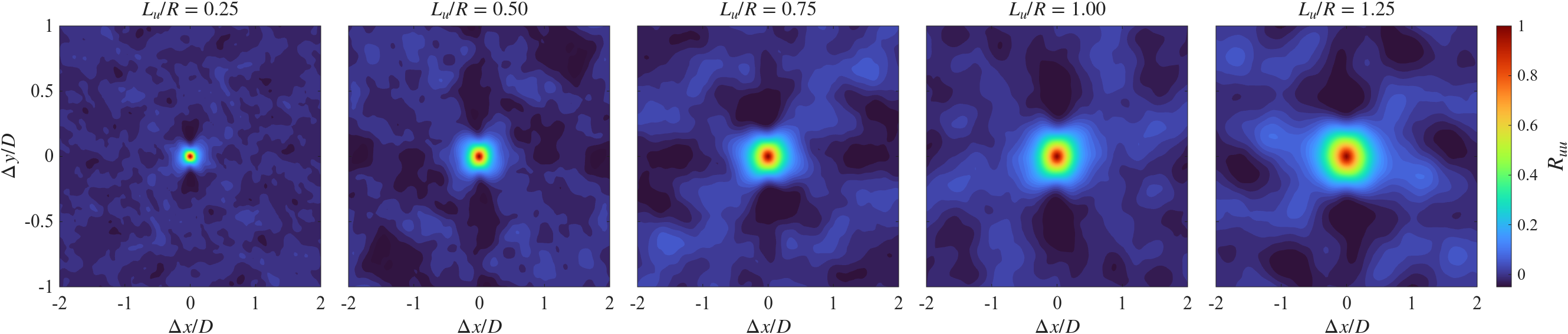}
        \caption{$R_{uu}$ correlation maps}
        \label{fig:Ruu_all}
    \end{subfigure}
    
    \vspace{1em}
    \begin{subfigure}[b]{\textwidth}
        \centering
        \includegraphics[width=\linewidth]{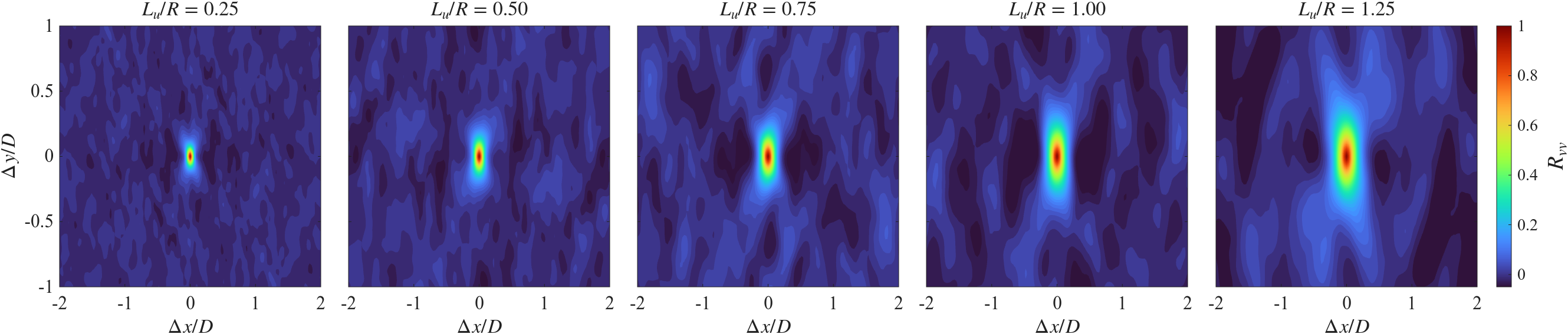}
        \caption{$R_{vv}$ correlation maps}
        \label{fig:Rvv_all}
    \end{subfigure}
    
    \vspace{0.5em}
    \begin{subfigure}[b]{\textwidth}
        \centering
        \includegraphics[width=\linewidth]{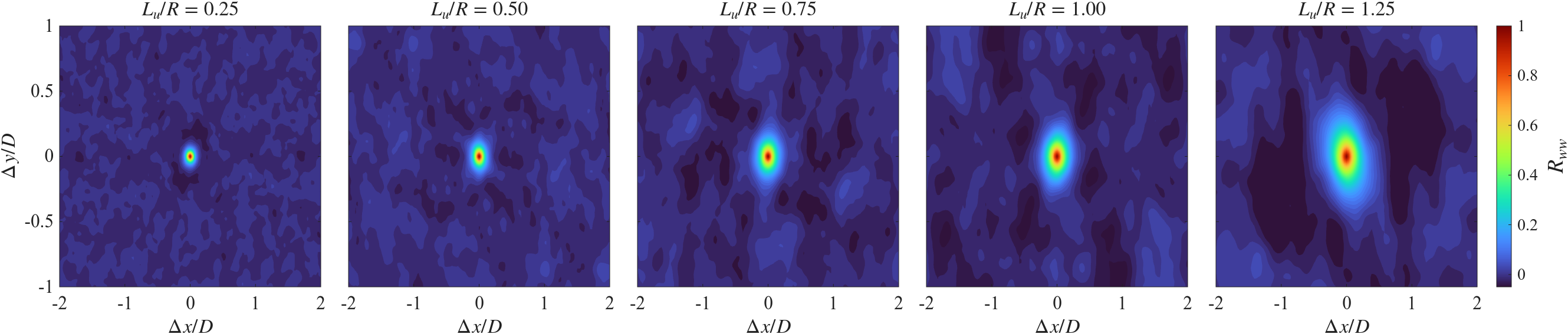}
        \caption{$R_{ww}$ correlation maps}
        \label{fig:Rww_all}
    \end{subfigure}

    \caption{
        Colormaps of the time‑averaged two‑point correlation functions of the streamwise ($R_{uu}$), lateral ($R_{vv}$), and vertical ($R_{ww}$) velocity components.  
        Each row shows correlations as functions of normalized streamwise and spanwise separations, $\Delta x/D$ and $\Delta y/D$, for several turbulence length scales $L_u/D$.  
        As $L_u/D$ increases, the correlation footprints broaden and change shape, reflecting the growth of large‑scale coherent structures in the flow.
    }
    \label{fig:R_corr_all}
\end{figure*}

The imposed inflow fields reproduce turbulent fluctuations with prescribed integral length scales $L_u/R$ ranging from $0.25$ to $1.25$ and without the existence of any turbine in the computational domain. The two-point correlation functions of the streamwise, lateral, and vertical velocity components ($R_{uu}$, $R_{vv}$, and $R_{ww}$, Figure~\ref{fig:R_corr_all}) measured at the first turbine location but with the absence of the turbine, provide a direct measure of the spatial coherence of the fluctuations. As $L_u/R$ increases, the correlation footprints broaden and retain significant values over longer separations, indicating the progressive growth of large-scale coherent eddies. For the smallest scales ($L_u/R=0.25$), the correlation rapidly decays within a fraction of the rotor diameter, yielding compact structures with limited streamwise extent. By contrast, the largest inflow scale ($L_u/R=1.25$) exhibits elongated streamwise correlations and a more isotropic footprint, consistent with the presence of long-wavelength energy-containing motions that can strongly modulate rotor-scale dynamics. The marked elongation of $R_{vv}$ relative to $R_{uu}$ highlights the anisotropy of cross-stream fluctuations, with lateral motions remaining more persistent in the streamwise direction.

\textcolor{black}{A quantitative summary of the flow turbulence characteristics is reported in Table~\ref{tab:inflow_stats}. All statistics are evaluated from time-resolved velocity signals sampled at a set of 13 probes distributed over the rotor-lateral and vertical axes, located at $x/D=0$ (corresponding to a distance of $6D$ downstream of the inlet), in the absence of the rotor. Across all cases, the mean streamwise velocity remains nearly constant ($\langle u \rangle \approx 11.3$–$11.4$~m/s), ensuring comparable mean inflow conditions. }

\begin{table*}
\centering
\caption{ Summary of flow turbulence characteristics at the rotor hub height ($x/D=0$), evaluated from the time series of the streamwise velocity component in the absence of the rotor for different inflow cases generated with varying turbulence integral length scales $L_u/R$. }
\label{tab:inflow_stats}
\begin{ruledtabular}
\begin{tabular}{@{}lrrrrrrrrr@{}}
Case & $\langle u \rangle$ [m/s] & $u'_{rms}$ [m/s] & $v'_{rms}$ [m/s] & $w'_{rms}$ [m/s] & $TI$ [-] & $f_{peak}$ [Hz] & $St$ [-] & ${v'}/{u'}$ & ${w'}/{u'}$ \\
\hline
$L_u/R=0.25$ &   11.41 &    0.241 & 0.204 & 0.202 & 0.019 & 0.064 & 0.711 & 0.847 & 0.838 \\

$L_u/R=0.50$ &   11.40 &    0.428 & 0.400 & 0.374 & 0.035 & 0.047 & 0.524 & 0.934 & 0.874 \\

$L_u/R=0.75$ &   11.39 & 0.515 & 0.493 & 0.472 & 0.044 & 0.027 & 0.300 & 0.963 & 0.958 \\

$L_u/R=1.00$ &   11.35 & 0.652 & 0.672 & 0.556 & 0.055 & 0.027 & 0.300 & 1.031 & 0.853 \\

$L_u/R=1.25$ &   11.27 & 0.930 & 0.793 & 0.783 & 0.072 & 0.011 & 0.122 & 0.853 & 0.842 \\
\end{tabular}
\end{ruledtabular}
\end{table*}

\begin{figure*}[ht]
  \centering
  \begin{subfigure}[t]{0.168\textwidth}
    \centering
    \includegraphics[width=\textwidth]{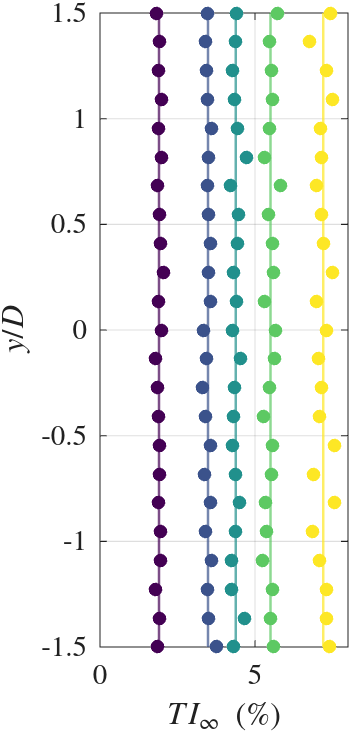}
    \caption{Lateral distribution of turbulence intensity.}
    \label{fig:ti-inflow}
  \end{subfigure}
  \hfill
  \begin{subfigure}[t]{0.395\textwidth}
    \centering
    \includegraphics[width=\textwidth]{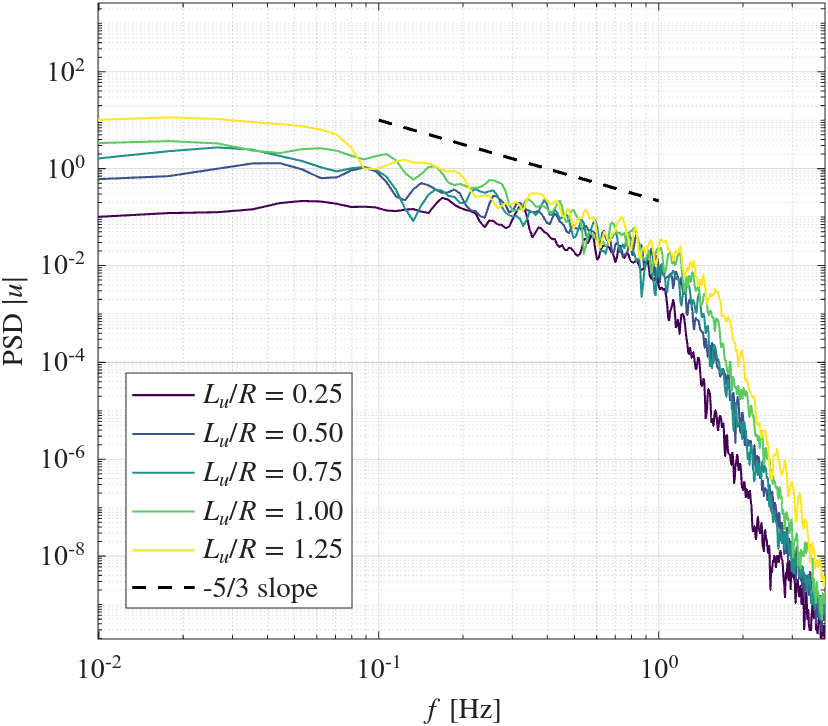}
    \caption{Velocity spectra with \(-5/3\) slope reference.}
    \label{fig:vel-psd}
  \end{subfigure}
  \hfill
  \begin{subfigure}[t]{0.395\textwidth}
    \centering
    \includegraphics[width=\textwidth]{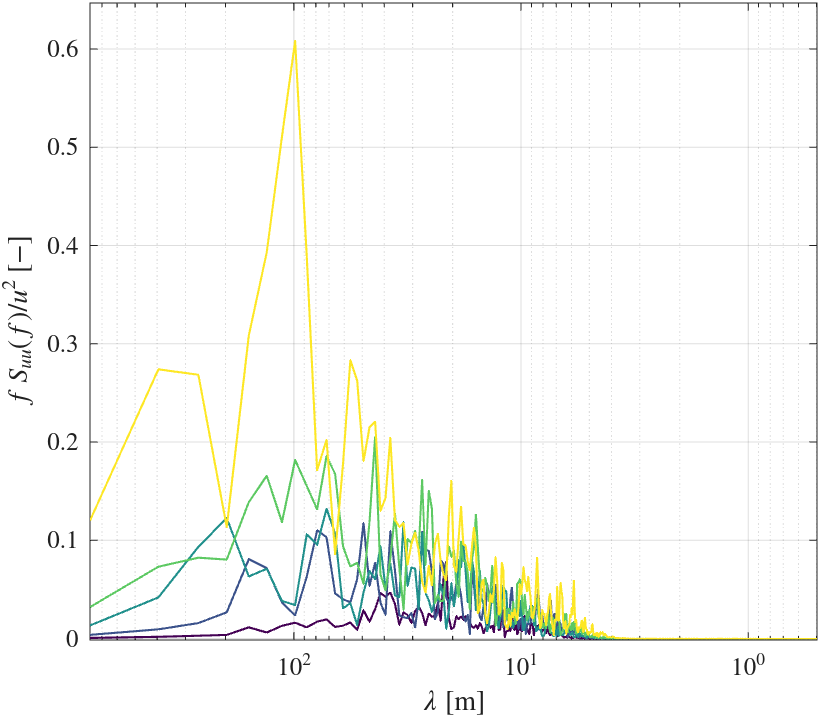}
    \caption{Pre-multiplied spectra as a function of wavelength.}
    \label{fig:premultiplied}
  \end{subfigure}

  \caption{Characterization of inflow turbulence for different integral length scales \(L_u\):
  (a) turbulence intensity distribution across the lateral direction,
  (b) velocity power spectral density showing inertial subrange scaling,
  and (c) pre-multiplied spectra highlighting energy distribution across wavelengths.}
  \label{fig:inflow-turbulence}
\end{figure*}

\textcolor{black}{
As the prescribed integral length scale at the computational domain inlet $L_u/R$ increases, the velocity fluctuations and the turbulence intensity at the measuring points increase systematically, with $TI$ rising from approximately $1.9\%$ at $L_u/R=0.25$ to $7.2\%$ at $L_u/R=1.25$. In fact, these results are quite understandable if we consider the information provided from equations \ref{turb_intensity} to \ref{two_point_correlation}, which shows the coupling between the length scale and turbulence intensity. Furthermore, these results agree very well with the observations based on the experimental tests made by references \cite{gambuzza2023influence} and \cite{bourhis2025impact}. 
}
The dominant frequency $f_{\text{peak}}$ is extracted directly from the power spectral density of the streamwise velocity time series, and the corresponding Strouhal number is defined as $St = f_{\text{peak}} D / U_\infty$. This definition reflects the dynamically realized temporal scales of the inflow turbulence, rather than Taylor’s frozen-flow hypothesis estimate based solely on the imposed integral length scale (i.e.\ $St \sim D/L_u$). With increasing $L_u/R$, the dominant frequency shifts toward lower values, leading to a systematic decrease of $St$ from approximately $0.71$ to $0.12$, indicating the emergence of larger and slower turbulent motions. The anisotropy ratios $v'/u'$ and $w'/u'$ remain close to unity across all cases, confirming that the synthetic inflow develops downstream while keeping a physical and realistic balance. Therefore, clarifying that the turbulence is intrinsically three-dimensional. 

\textcolor{black}{
The lateral distribution of turbulence intensity (Figure~\ref{fig:ti-inflow}), evaluated along the $y$ direction at $x/D=0$ and $z/D=0$, demonstrates that the turbines are exposed to homogeneous conditions across the swept area.}
Spectral analysis of the streamwise velocity (Figure~\ref{fig:vel-psd}) further illustrates the influence of $L_u/R$ on the energetic content of the inflow. All cases reproduce a clear inertial sub-range, following the Kolmogorov $-5/3$ slope, which attests to the physical realism of the synthetic inflow. 
As mentioned in Table \ref{tab:inflow_stats}, increasing $L_u/R$ shifts the spectral peak toward lower frequencies and enhances the energy at large scales, which corresponds to the stronger spatial coherence observed in the correlation maps. In non-dimensional form, the Strouhal numbers associated with the spectral peak decrease from $St \approx 0.71$ ($L_u/R=0.25$) to $St \approx 0.12$ ($L_u/R=1.25$), evidencing a systematic modulation of the dominant turbulent time scales.
In addition, Figure~\ref{fig:premultiplied} presents the pre-multiplied spectra
plotted against the streamwise wavelength ($\lambda = U_\infty/f$). In this representation, the area under each curve corresponds directly to the turbulent kinetic energy. The results clearly show that cases with larger $L_u$ contain energy across a much broader range of wavelengths, with distinct peaks extending beyond the rotor radius but not reaching the full rotor diameter. By contrast, the smallest integral scale case ($L_u/R=0.25$) exhibits a significantly narrower distribution, confirming that only a limited subset of turbulent scales is active.

Overall, these inflow cases span a broad range of turbulence length scales while maintaining consistent mean conditions. They thus provide a controlled framework to examine how the scale of ambient turbulence governs wake development, coherent structure persistence, and subsequent turbine–wake interactions.

\subsection{Wake Structure and Recovery}

\begin{figure*}
  \centering
  \includegraphics[width=\textwidth]{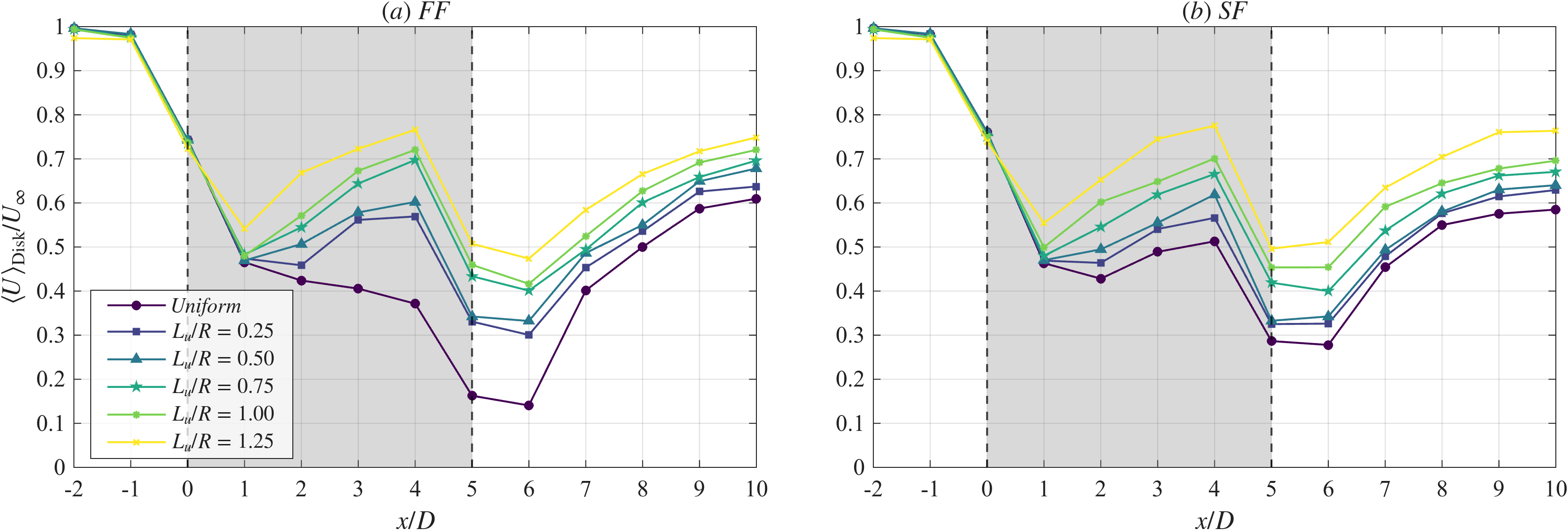}
  \caption{Disk-averaged, streamwise velocity normalized by the freestream,
  \(\langle U\rangle_{\text{Disk}}/U_\infty\), versus axial position \(x/D\) for two rotor-mount configurations: (a) FF (fixed–fixed) and (b) SF (surge–fixed). Vertical black dashed lines mark the rotor locations; the gray shaded region indicates the wake between the rotors. 
  The legend shows the inflow cases with different integral length scales \(L_u\).}
  \label{fig:Udisk-FF-SF}
\end{figure*}

\begin{figure*}
    \centering
    \begin{subfigure}[b]{0.49\textwidth}
        \centering
        \includegraphics[width=\linewidth]{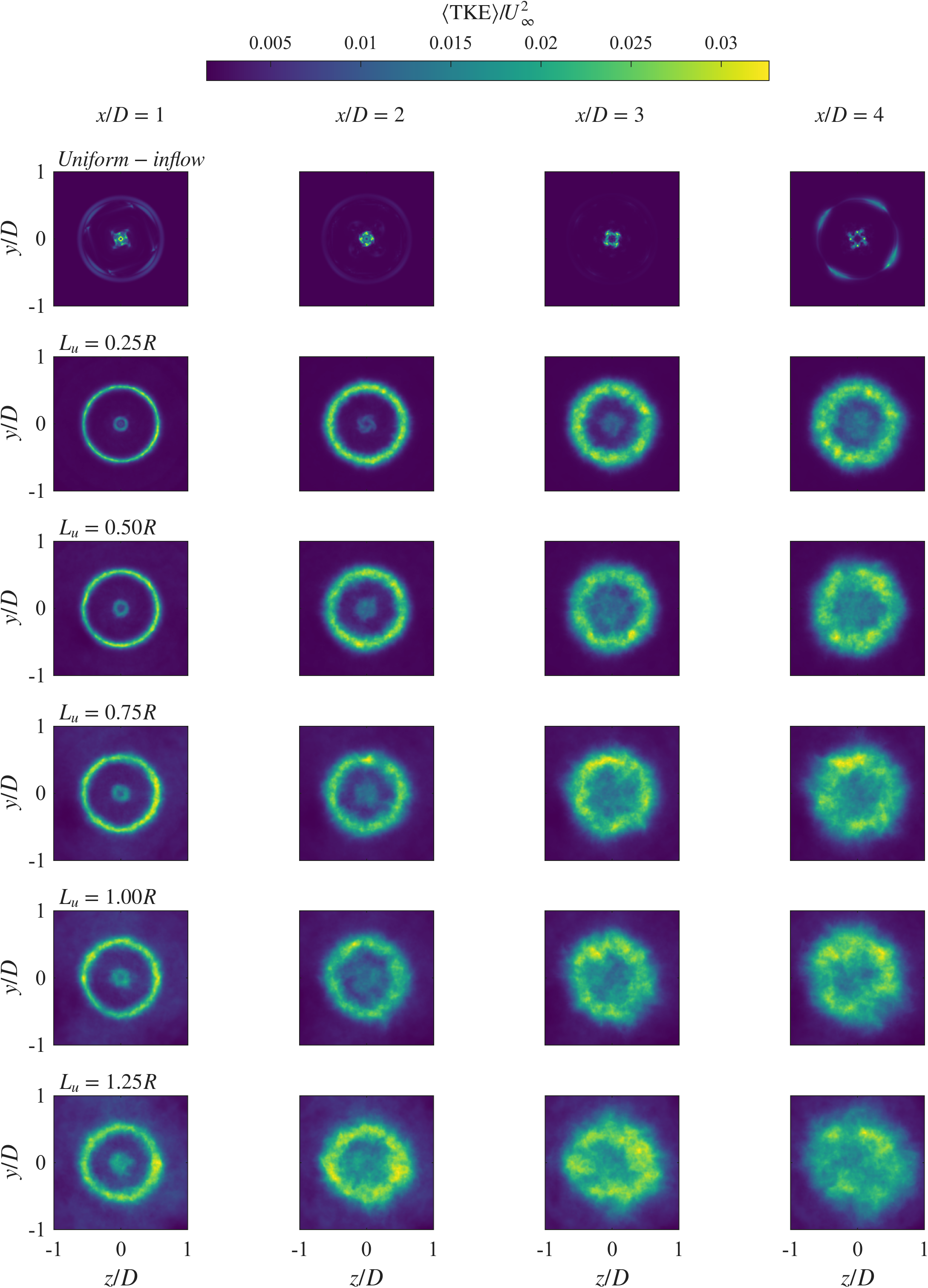}
        \caption{FF configuration (both fixed).}
        \label{fig:TKE_FF}
    \end{subfigure}
    \hfill
    \begin{subfigure}[b]{0.49\textwidth}
        \centering
        \includegraphics[width=\linewidth]{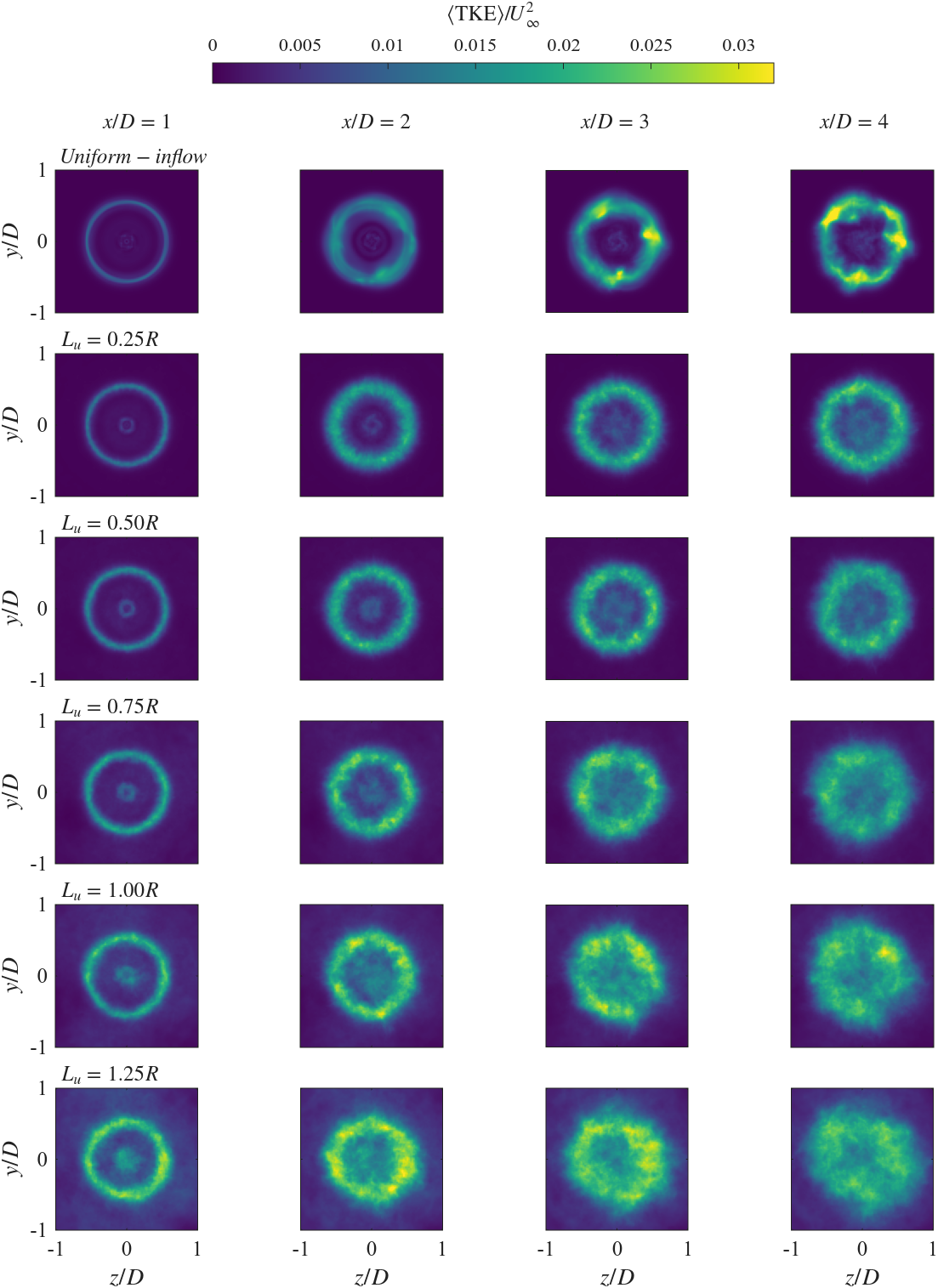}
        \caption{SF configuration (surging–fixed).}
        \label{fig:TKE_SF}
    \end{subfigure}

    \caption{
        Normalized mean turbulent kinetic energy, \(\langle \mathrm{TKE} \rangle / U_\infty^2\), in cross‑wake planes at downstream positions \(x/D = 1, 2, 3, 4\).  
        Each montage shows the effect of a uniform inflow (top row) and successively increasing turbulence length scales \(L_u\) (rows 2–6) on the wake structure.  
        (a) Surging–fixed (SF) case, where the downstream rotor surges and the upstream rotor is fixed; (b) fixed–fixed (FF) case, with both rotors fixed.
    }
    \label{fig:TKE_SF_FF}
\end{figure*}

\begin{figure*}
  \centering
  \begin{subfigure}[t]{0.48\textwidth}
    \centering
    \includegraphics[width=\textwidth]{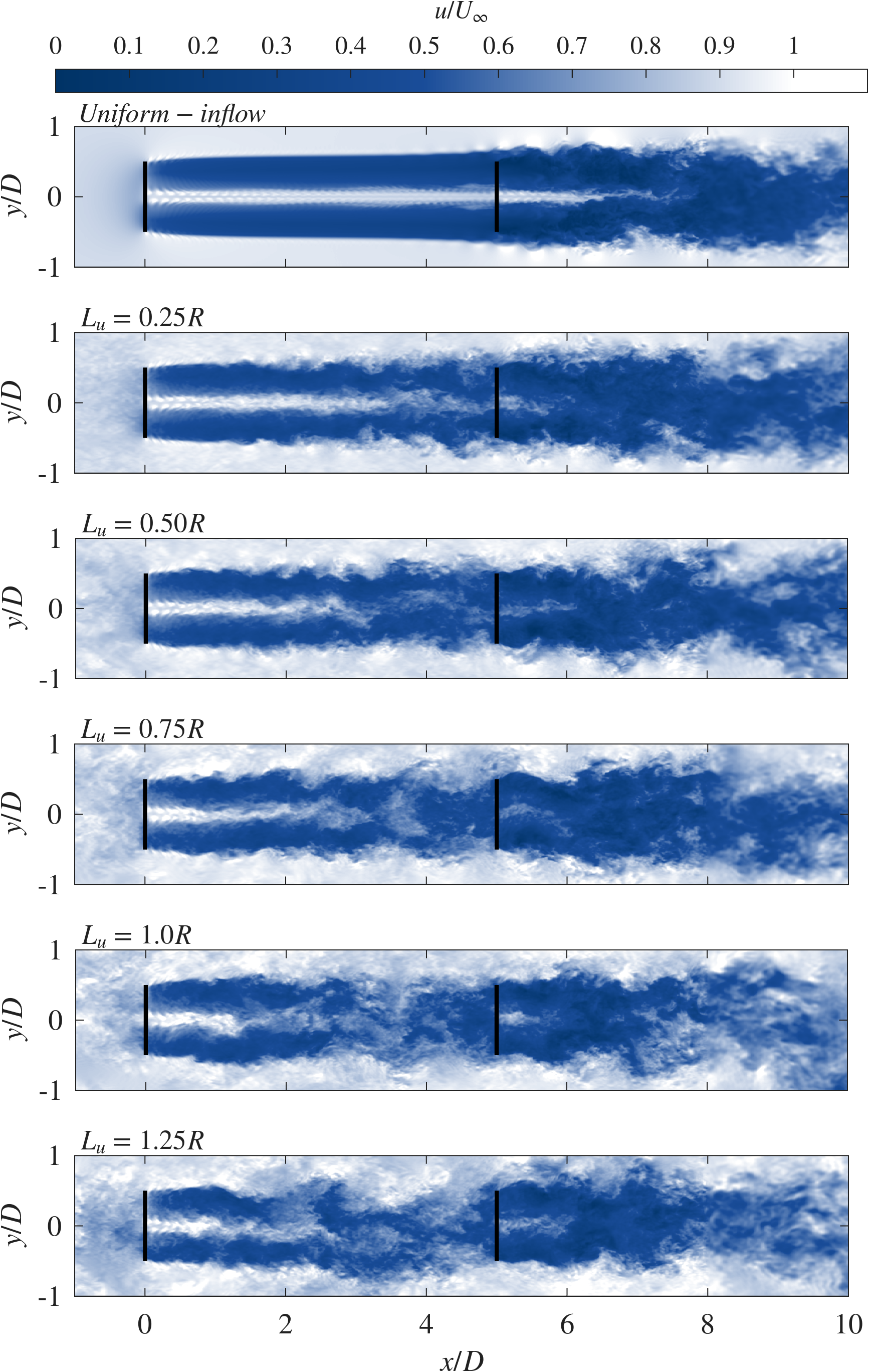}
    \caption{FF configuration (both fixed).}
    \label{fig:velocity-ff}
  \end{subfigure}
  \hfill
  \begin{subfigure}[t]{0.48\textwidth}
    \centering
    \includegraphics[width=\textwidth]{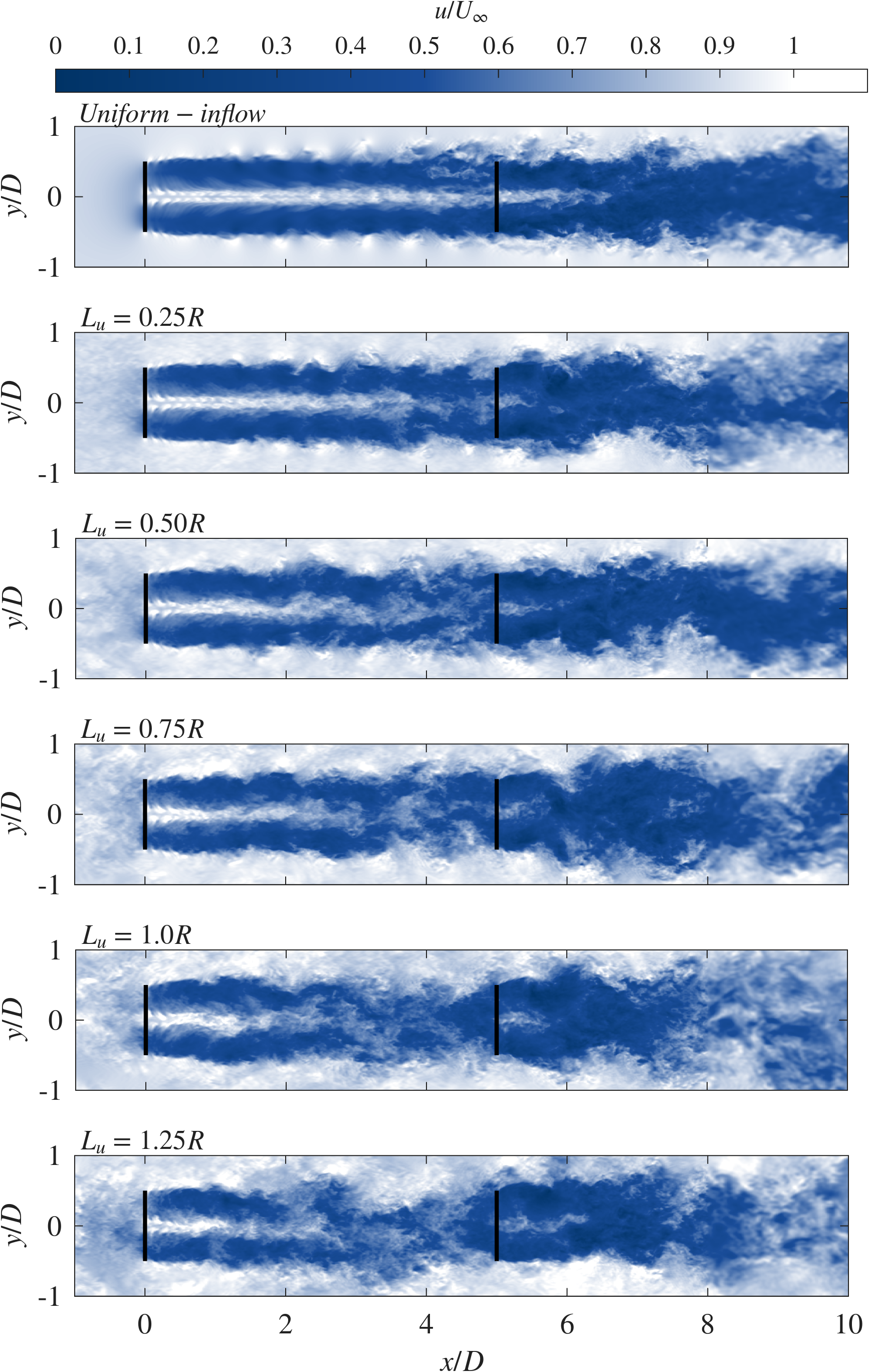}
    \caption{SF configuration (surging–fixed).}
    \label{fig:velocity-sf}
  \end{subfigure}

  \caption{Normalized instantaneous velocity fields in the \(x\)–\(y\) plane for different integral length scales \(L_u\). Results are shown for (a) fix–fix and (b) surg–fix rotor configurations. The black vertical lines at \(x/D = 0\) and \(x/D = 5\) indicate the rotors locations.}
  \label{fig:velocity-instantaneous}
\end{figure*}

Figure~\ref{fig:Udisk-FF-SF} presents the non-dimensional disk-averaged streamwise velocity, $\langle U\rangle_{\text{Disk}}/U_\infty$, along the centerline for the two rotor configurations studied: (a) fixed–fixed (FF) and (b) surge–fixed (SF). The vertical dashed lines indicate the rotor planes at $x/D=0$ and $x/D=5$, while the shaded region marks the inter-turbine wake.

Immediately downstream of the upstream rotor ($x/D \lesssim 1$), all cases experience a strong velocity drop, followed by partial recovery as the wake begins to mix with the surrounding flow. How much recovery occurs before the downstream turbine depends strongly on both the inflow turbulence length scale and the motion of the upstream platform. Under uniform inflow, the FF case maintains a pronounced velocity deficit, with $\langle U\rangle_{\text{Disk}}/U_\infty \approx 0.37$ at $x/D=4$, corresponding to a $63\%$ reduction relative to the freestream. When the upstream turbine is allowed to surge, this deficit is reduced to about $49\%$ at the same location ($\langle U\rangle_{\text{Disk}}/U_\infty \approx 0.51$), indicating that platform motion enhances mixing within the inter-turbine region.

Increasing the inlet turbulence integral length scale $L_u/R$ further accelerates wake recovery in both configurations. As $L_u/R$ increases, large incoming eddies interact more effectively with the rotor-generated shear layers, leading to stronger mixing and a more uniform velocity distribution between the turbines. At $x/D=4$, $\langle U\rangle_{\text{Disk}}/U_\infty$ increases monotonically with $L_u/R$, and for the largest value considered ($L_u/R=1.25$), the velocity deficit falls below approximately $24\%$ in the FF case and $22\%$ in the SF case. This behavior is consistent with the inflow characteristics discussed earlier: larger $L_u/R$ shifts energy toward lower frequencies and increases spatial coherence, allowing more efficient momentum transfer into the wake.

A second velocity minimum appears downstream of the fixed downstream turbine ($x/D \approx 5$–6). For the FF configuration under uniform inflow, this deficit is particularly deep, but the wake recovers rapidly farther downstream, reaching a more uniform profile by $x/D \approx 10$. In the SF configuration, the post-rotor deficit is consistently shallower, and the recovery occurs more smoothly for all values of $L_u/R$. The surge motion of the upstream turbine introduces time-dependent disturbances that thicken the shear layer and promote large-scale unsteady structures (see Figures~\ref{fig:q_iso_surg} and~\ref{fig:p_iso_surg}), which enhance momentum exchange across the wake. As a result, the SF curves remain above their FF counterparts over $6 \lesssim x/D \lesssim 10$, indicating universal improvement in wake recovery.

Two distinct regimes can be identified. For uniform inflow and small turbulence length scales ($L_u/R \lesssim 0.5$), the background turbulence contributes little to mixing, and the relative benefit of surge motion is therefore most pronounced. In contrast, for larger turbulence scales ($L_u/R \gtrsim 0.75$), both FF and SF wakes recover more rapidly due to strong inflow-driven mixing, and the additional benefit of surge becomes smaller in relative terms. In all cases, increasing $L_u/R$ and introducing upstream surge shift the wake toward earlier breakdown and faster re-energization both between and beyond the turbines.

Overall, the axial velocity profiles show that increasing the inflow integral length scale systematically weakens the inter-turbine velocity deficit and accelerates wake recovery. Superimposed surge motion of the upstream turbine further enhances this process, reducing the inter-turbine deficit by up to about 14 percentage points at $x/D=4$ under uniform inflow (from $63\%$ to $49\%$) and leading to consistently shallower velocity deficits downstream of the second turbine across all turbulence conditions.

Figure~\ref{fig:TKE_SF_FF} presents cross-wake sections of the normalized mean turbulent kinetic energy (total), $\langle \mathrm{TKE} \rangle / U_\infty^2$, at downstream locations $x/D=1, 2, 3,$ and $4$ for the (a) fixed–fixed (FF) and (b) surge–fixed (SF) configurations, under uniform inflow and five turbulent inflows with increasing integral length scale $L_u/R$. These maps illustrate how the shear layer generated by the rotor tip vortices forms and how its subsequent breakdown depends on both the inflow turbulence scale and the upstream platform motion.

Under uniform inflow, the FF configuration exhibits a narrow annular region of elevated TKE at $x/D=1$, corresponding to the tip-shear layer surrounding a relatively quiescent wake core. This annular structure remains thin and well organized up to $x/D \approx 3$, indicating limited radial transport of turbulence into the wake interior. Even at $x/D=4$, the ring structure persists, reflecting delayed shear-layer breakdown and continued shielding of the wake core. In contrast, for the SF configuration, the same annular feature forms near the rotor but rapidly thickens and becomes azimuthally non-uniform. At $x/D=2$–3, the ring fragments and elevated TKE fill the wake core, indicating earlier shear-layer destabilization, which leads to enhanced entrainment. This behavior is consistent with the axial velocity profiles shown in Figure~\ref{fig:Udisk-FF-SF}, where surge motion leads to a reduced inter-turbine velocity deficit.

For both configurations, increasing the inlet turbulence length scale $L_u/R$ significantly accelerates the breakdown of the tip-shear layer. Larger $L_u/R$ introduces energetic, low-frequency eddies that interact strongly with the rotor-generated vortices, causing the annular TKE structure to broaden and lose coherence already at $x/D=1$–2. At $x/D=3$, the ring is largely disrupted, and the wake core contains substantial turbulent kinetic energy. This evolution is monotonic across the rows of Figure~\ref{fig:TKE_SF_FF}: from a compact, well-defined ring with limited core activity at $L_u/R=0.25$ to a diffuse shear layer and a highly energized wake core at $L_u/R=1.25$. These trends are consistent with the inflow spectral and two-point correlation characteristics discussed earlier, as well as with the higher inter-turbine disk-averaged velocities shown in Figure~\ref{fig:Udisk-FF-SF}.

At a given inflow turbulence length scale, the SF configuration consistently exhibits (i) a thicker annular region of elevated TKE near $x/D=1$, (ii) earlier azimuthal breakup of the shear layer and penetration of turbulence into the wake core by $x/D=2$–3, and (iii) a more uniform TKE distribution across the wake cross-section at $x/D=4$. The upstream surge motion periodically modulates the shear layer and promotes large-scale unsteadiness, which enhances Reynolds-stress gradients across the wake boundary. This leads to an earlier collapse of the organized tip-vortex structure and faster re-energization of the wake, explaining the reduced velocity deficits observed upstream of the downstream rotor.

Taken together, the TKE distributions demonstrate that both increasing the inflow turbulence integral length scale and allowing upstream surge motion, promote earlier shear-layer breakdown and more efficient mixing within the wake. These mechanisms shorten the near-wake region, weaken the inter-turbine velocity deficit, and create more favorable inflow conditions for the downstream turbine, which is consistent with the performance trends discussed later.

To complement the averaged velocity and TKE statistics, Figure~\ref{fig:velocity-instantaneous} presents snapshots of the instantaneous normalized streamwise velocity in the $x$–$y$ plane for the FF and SF configurations. These instantaneous fields provide direct visual evidence of the flow structures responsible for the mean wake behavior discussed above.

As shown in Figure~\ref{fig:velocity-instantaneous}, for the FF configuration under uniform inflow, the wake is characterized by a highly ordered structure, with two nearly symmetric shear layers enclosing a persistent low-velocity core. The coherence and symmetry of these shear layers explain the deep velocity deficit and slow recovery observed in the disk-averaged profiles (Figure~\ref{fig:Udisk-FF-SF}). When the upstream turbine surges (SF), this symmetry is visibly broken: the shear layers exhibit pronounced lateral displacements and intermittent thickening, leading to a more diffuse and irregular wake footprint.

With increasing inflow turbulence length scale, the instantaneous wake structure becomes progressively less organized. Rather than forming smooth, continuous shear layers, the wake exhibits strong spatial variability, with intermittent high- and low-speed patches extending across the wake cross-section. These instantaneous distortions correspond to the earlier breakdown of the shear layer and enhanced core mixing identified in the TKE maps (Figure~\ref{fig:TKE_SF_FF}), and they explain the faster recovery of the mean velocity deficit observed in Figure~\ref{fig:Udisk-FF-SF}. The surge motion introduces time-dependent perturbations that continually disturb the shear layer, preventing the formation of long-lived coherent structures. As a result, momentum exchange with the surrounding flow is enhanced, consistent with the reduced inter-turbine velocity deficits seen in the time-averaged results.

The instantaneous velocity fields thus provide a physical bridge between the statistical measures and the underlying wake dynamics. They show how coherent shear layers persist under fixed, low-disturbance conditions, while unsteadiness introduced by flow turbulence and upstream motion destabilizes these structures and accelerates wake recovery upstream of the downstream turbine.

\subsection{Wake Coherent Structures and Spectral Characteristics}

\begin{figure*}
    \centering
    \subfloat[FF configuration (both fixed).]{%
        \includegraphics[width=0.48\textwidth]{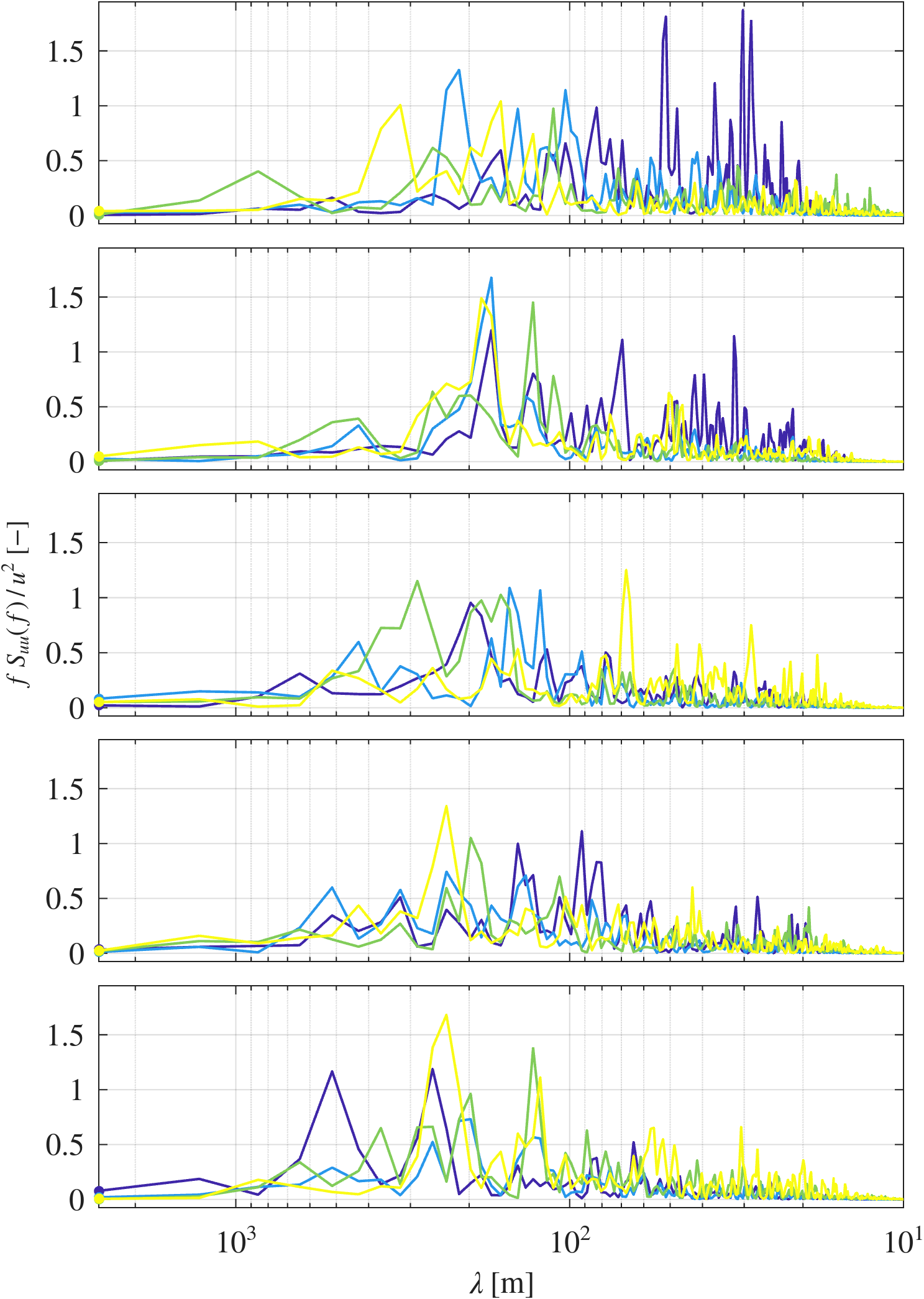}
    }
    \hfill
    \subfloat[SF configuration (surging–fixed).]{%
        \includegraphics[width=0.48\textwidth]{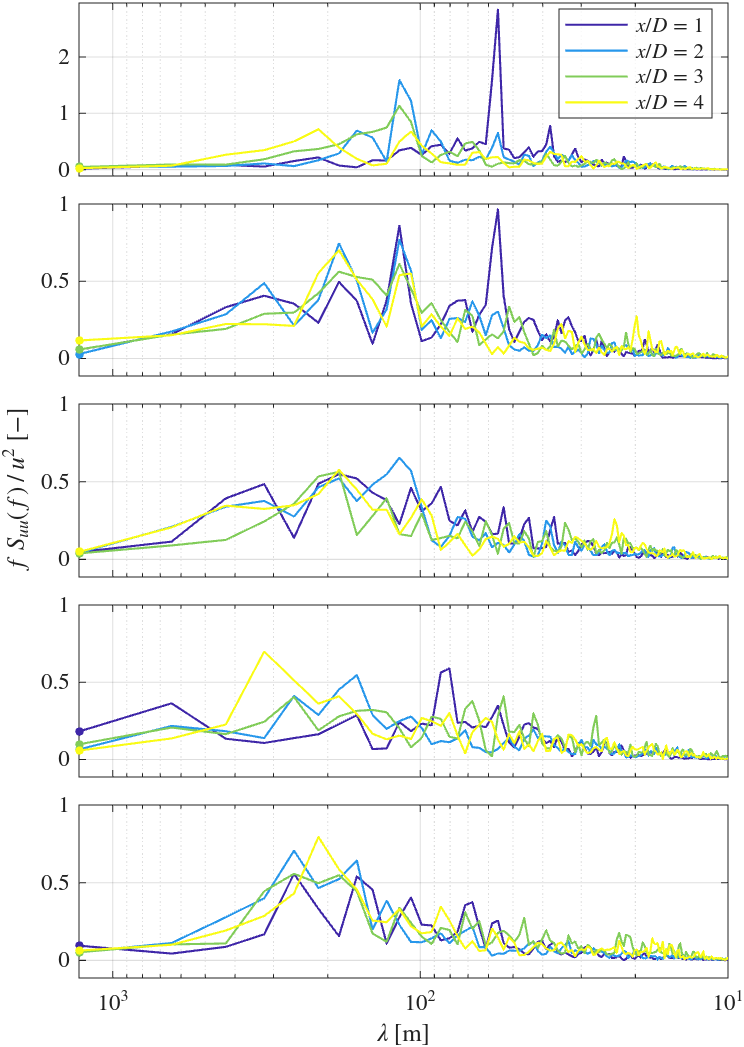}
    }
    \caption{
        Pre-multiplied power spectral density $fS_{uu}(f)/u^2$ of streamwise velocity fluctuations at various wake locations ($x/D = 1$, 2, 3, 4). 
        Each row corresponds to a different turbulence length scale: 
        from top to bottom, $L_u/R = 0.25$, 0.50, 0.75, 1.00, and 1.25. 
        (a) Fixed–Fixed configuration; 
        (b) Surge–fixed configuration. 
        Spectra are plotted versus wavelength $\lambda = U_\infty / f$ in log scale to highlight the shift in energy distribution across wake regions and turbulence scales.
    }
    \label{fig:premult-spectra}
\end{figure*}

\begin{figure*}
    \centering
    \includegraphics[width=0.9\linewidth]{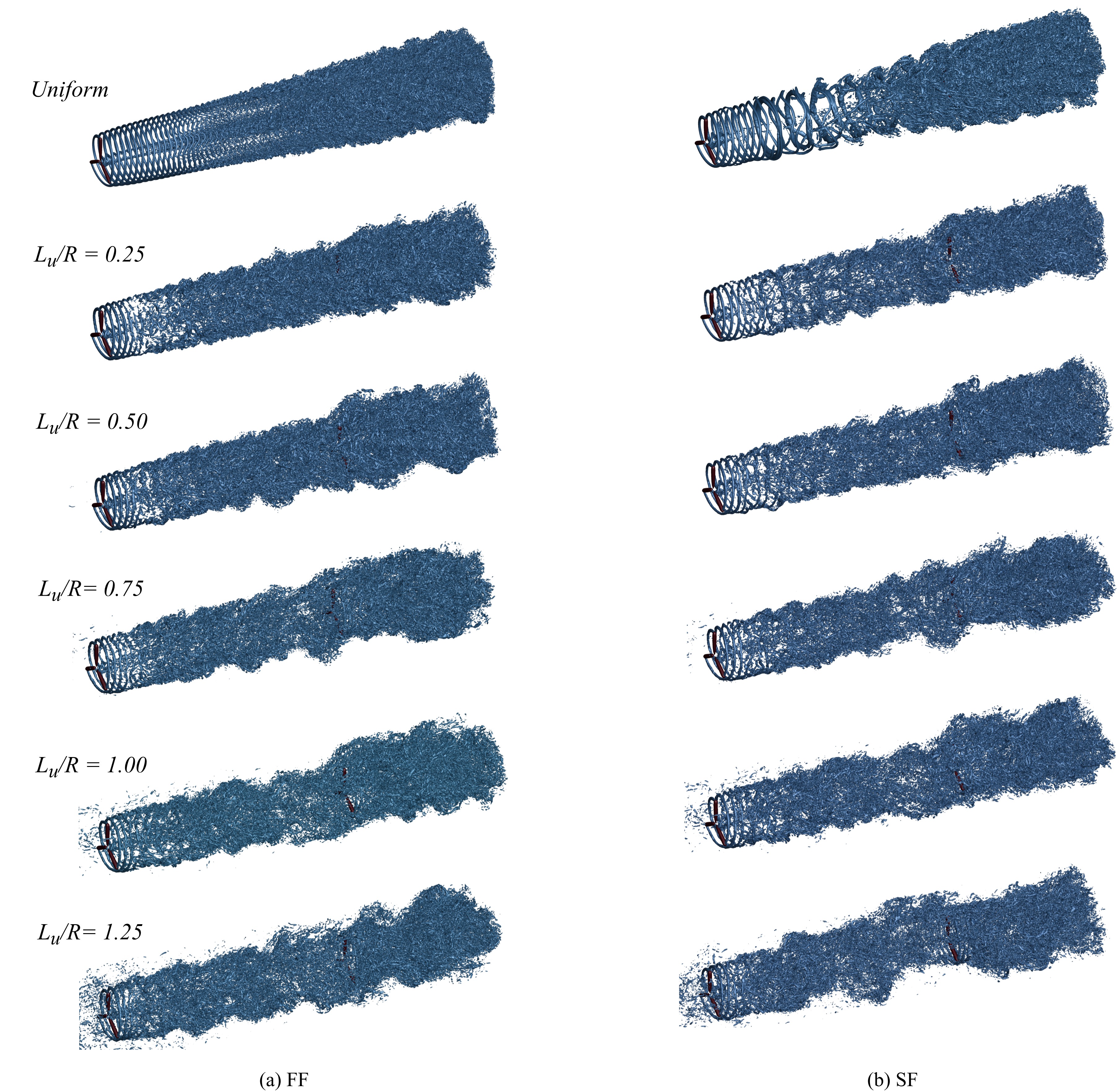}
  \caption{
    Instantaneous vortical structures visualized by the $Q$-criterion ($Q=0.25s^{-2}$) for 
    different inflow turbulence scales $L_u/R$. 
    Iso surfaces illustrate the evolution of tip-vortex systems and their breakdown in (a) the fixed–fixed (FF) and (b) surge–fixed (SF) configurations. 
  }
    \label{fig:Qiso_all}
\end{figure*}

Figure~\ref{fig:premult-spectra} introduces the pre-multiplied power spectral density, $fS_{uu}(f)/u^{2}$, of the streamwise velocity fluctuations at different downstream positions ($x/D = 1, 2, 3, 4$). At each $x/D$ the same 13 probes defined in section \ref{TI-initial-result} were employed to obtain the data presented in 
Figure~\ref{fig:premult-spectra}. Each row corresponds to a distinct inflow turbulence length scale $L_u/R$, while panels (a) and (b) represent the FF and SF configurations, respectively. The spectra are plotted against the streamwise wavelength ($\lambda = U_\infty/f$) to emphasize the redistribution of turbulent kinetic energy across spatial scales as the wake evolves.

Just downstream of the upstream rotor, at ($x/D = 1$) and particularly when $L_u/R=0.25$, the energy is concentrated at wavelengths of the order of the rotor radius, associated with the coherent tip-vortex system. Note also that the effect of upstream surge motion organizes the tip vortical structures into predominant wavelengths having a considerable energy level. As we move downstream ($x/D = 2, 3, 4$), the spectral peak progressively broadens and shifts toward longer wavelengths (with an associated smaller spectral density), reflecting the roll-up and merging of tip vortices into larger-scale, meandering motions. 
These findings align very well with the observations made by \cite{vahidi2024influence}. 

When increasing the inflow turbulence length scale to $L_u/R=0.5$ (the second row of Figure~\ref{fig:premult-spectra}), particularly for the SF configuration at $x/D=1$, a new peak having a higher wavelength is observed. This second peak is likely to characterize the mixing of the tip helical vortical structures observed in Figure~\ref{fig:Qiso_all}b. When moving downstream, the peaks broaden and displace to longer wavelengths as previously reported. 
At larger inflow length scales and regardless of the $x/D$ position considered, peaks become wider, characterizing the different frequencies associated with the different scale flow structures involved in the flow mixing process. This process is particularly enhanced by the upstream turbine surge displacement.

Compared to the FF case, the SF configuration exhibits 
a more rapid broadening of the spectra with downstream distance. The surge motion introduces additional unsteadiness that interacts with the incoming flow and promotes mixing.
This behavior is consistent with the reduced velocity deficits observed in Figure~\ref{fig:Udisk-FF-SF} and the elevated TKE levels in the wake core. 

The pre-multiplied spectra quantitatively confirm that both the inflow turbulence scale and surge motion broaden the range of energetic wavelengths within the wake. The area under each curve (proportional to the turbulent kinetic energy) increases and shifts toward longer wavelengths for larger $L_u$, demonstrating that the wake becomes dominated by low-frequency, large-scale motions as it recovers downstream.

Figure~\ref{fig:Qiso_all} illustrates the instantaneous vortical structures identified by the $Q$-criterion for both (a) FF and (b) SF configurations under different inflow turbulence scales $L_u/R$. The iso surfaces capture the evolution of the tip-vortex system and its subsequent breakdown along the wake.

For the uniform inflow, the FF case displays a highly ordered helical lattice of tip vortices extending several rotor diameters downstream. The wake core remains uniform-like, with limited cross-shear interaction between successive vortex turns. This persistence of coherent helices corresponds to the low TKE core and narrow spectral energy range observed earlier. In contrast, the SF case already shows visible deformation of the helical structure: surge motion induces periodic modulation of the tip-vortex spacing and amplitude, introducing large-scale oscillations that accelerate mutual induction and vortex merging.

Increasing $L_u/R$ progressively destabilizes the helical pattern. For $L_u/R=0.25$, the tip vortices remain well defined but begin to exhibit mild waviness; by $L_u/R=0.75$, the structure transitions into fragmented vortex bundles, and for $L_u/R\ge1.0$ the coherent helices are largely destroyed, replaced by a turbulent cloud of small-scale structures. This evolution mirrors the broadening of the pre-multiplied spectra and the enhanced TKE levels in the near-wake cross sections. The larger-scale inflow eddies provide the low-frequency perturbations necessary to excite shear-layer instabilities and trigger earlier vortex breakdown.

Across all values of $L_u/R$, the SF configuration shows earlier and more pronounced breakdown of the tip-vortex system than the FF case. The periodic surge motion continuously disturbs the tip-shear layer, enhancing vortex interaction and lateral motion. As a result, the SF wakes exhibit thicker shear regions and faster recovery of streamwise momentum, consistent with the velocity profiles and TKE distributions shown in Figure~\ref{fig:TKE_SF_FF}. These flow features indicate that platform motion reinforces the mixing already introduced by large-scale inflow turbulence, leading to a shorter and more rapidly recovering near wake.

\subsection{Surging Phase Influence}

In this subsection, both turbines are allowed to surge, and the influence of the relative phase between their motions is examined. Two-phase configurations are considered: in-phase motion, where the upstream and downstream turbines surge synchronously ($\Delta \phi_S = 0$), and anti-phase motion, where they surge with a phase difference of $\pi$ ($\Delta \phi_S = \pi$), see Figure~\ref{fig:phase_motion}. These cases are analyzed under uniform inflow conditions and under turbulent inflow with the largest integral length scale considered in this study ($L_u/R = 1.25$). 

\begin{figure*}[ht!]
    \centering
    \subfloat[Uniform inflow.]{%
        \includegraphics[width=0.49\textwidth]{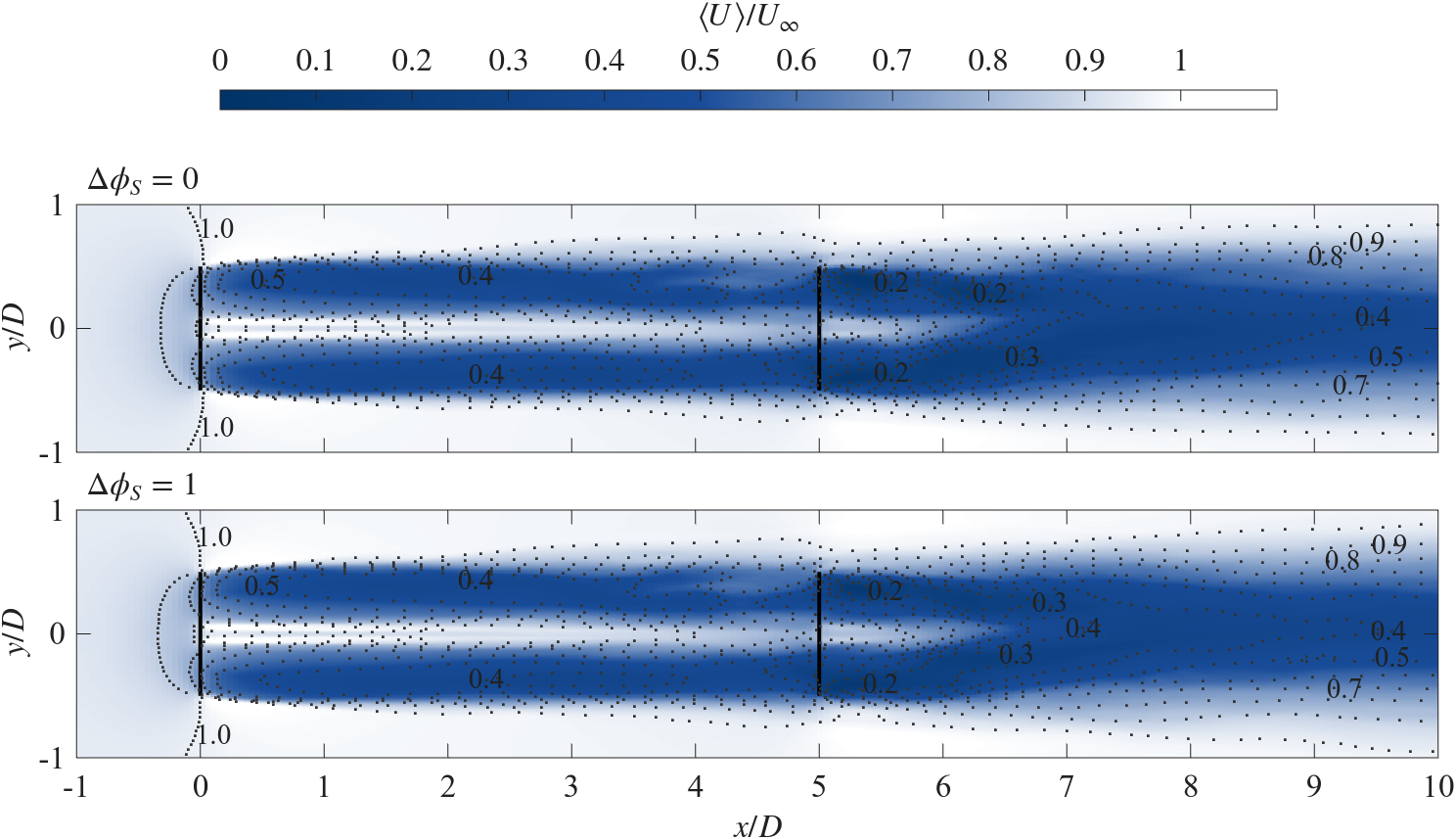}
    }
    \hfill
    \subfloat[Turbulent inflow ($L_u/R = 1.25$).]{%
        \includegraphics[width=0.49\textwidth]{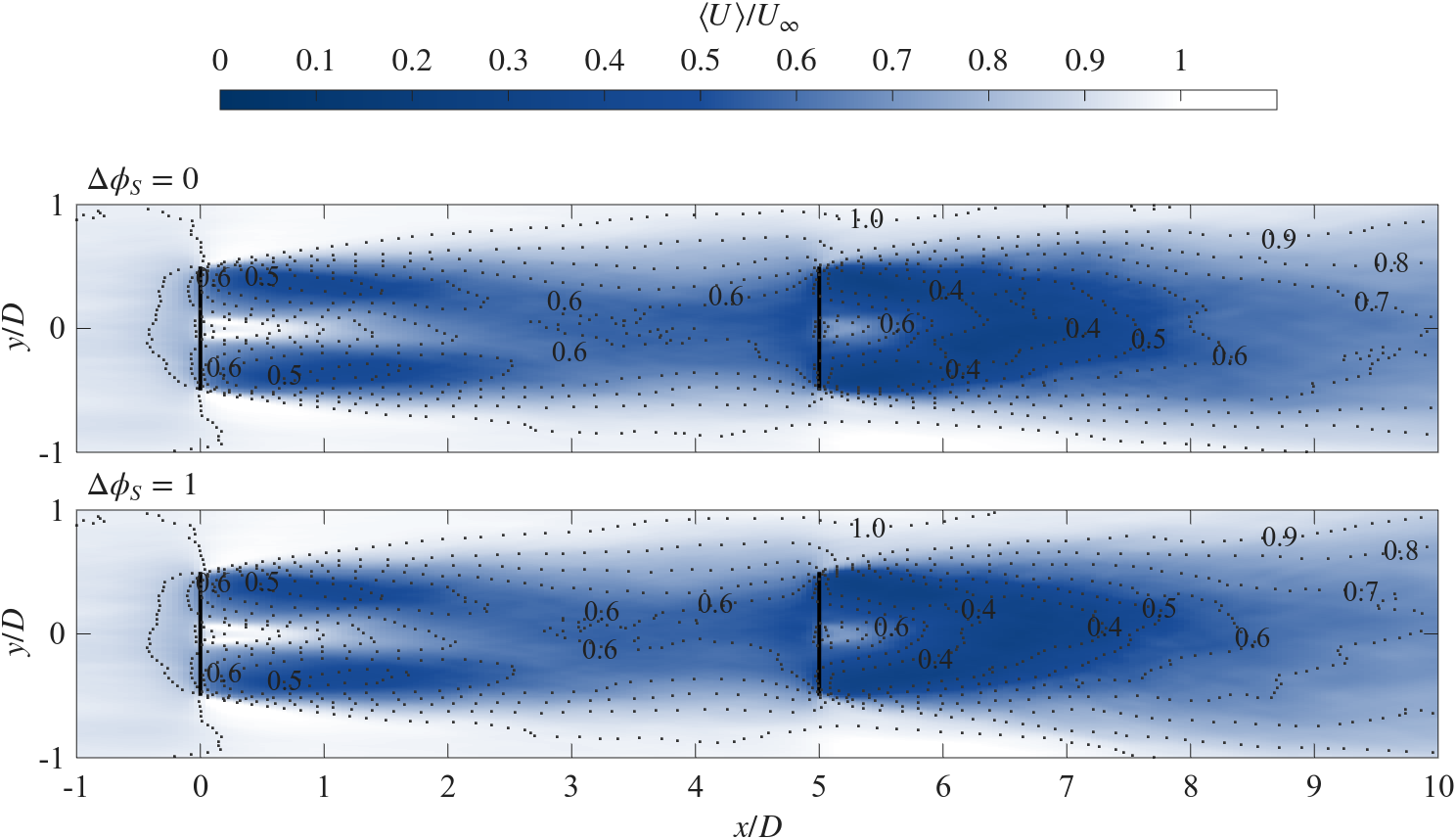}
    }
    \caption{
        Time-averaged normalized streamwise velocity fields $U/U_\infty$ for surging-phase configurations of the upstream turbine.
        The left panel corresponds to uniform inflow conditions, while the right panel shows the turbulent inflow case with the largest integral length scale, $L_u/R = 1.25$.
        In each panel, the top row represents in-phase surging motion ($\Delta\phi_S = 0$), and the bottom row represents anti-phase motion ($\Delta\phi_S = \pi$).
    }
    \label{fig:SP_uniform}
\end{figure*}

\begin{figure*}[ht!]
    \centering
    \subfloat[Uniform inflow.]{%
        \includegraphics[width=0.49\textwidth]{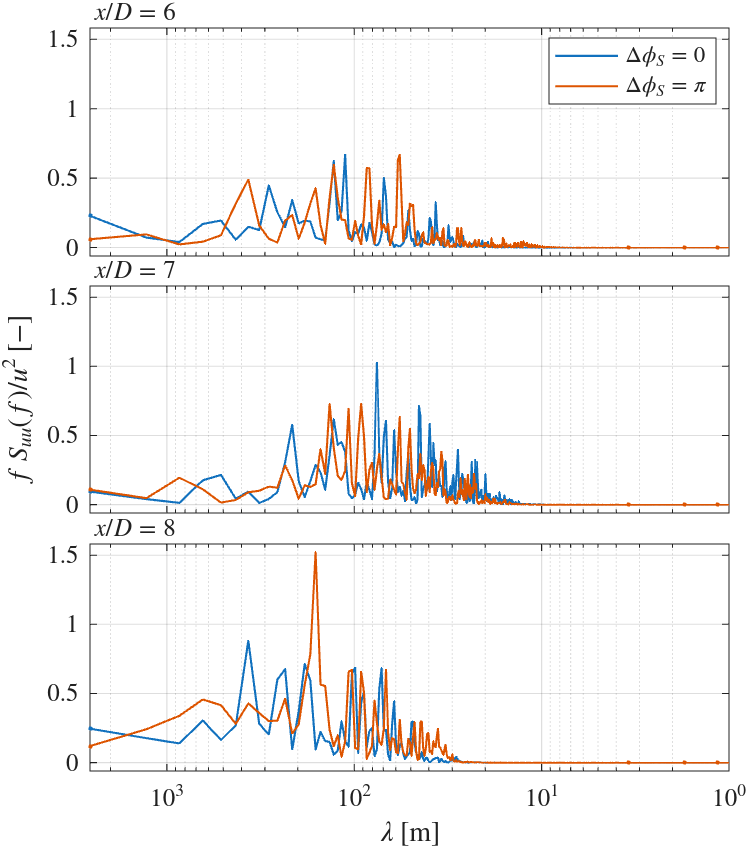}
    }
    \hfill
    \subfloat[Turbulent inflow ($L_u/R = 1.25$).]{%
        \includegraphics[width=0.49\textwidth]{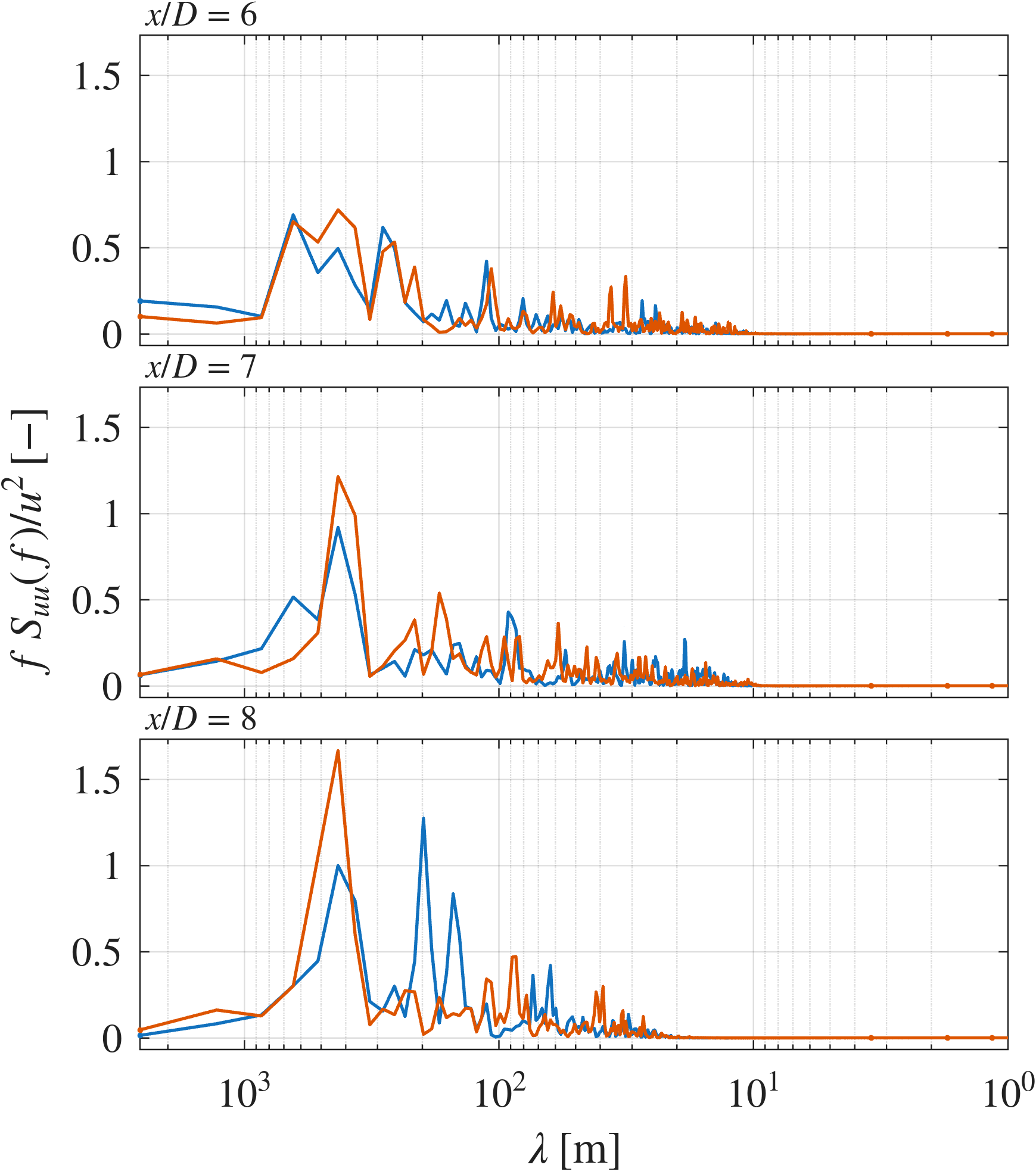}
    }
    \caption{
        Normalized pre-multiplied power spectral density $f S_{uu}(f)/u'^2$ of streamwise velocity fluctuations plotted versus wavelength $\lambda = U_\infty / f$ for downstream wake locations $x/D = 6$, 7, and 8.
        The left panel corresponds to uniform inflow conditions, while the right panel shows the turbulent inflow case with the largest integral length scale, $L_u/R = 1.25$.
        In each panel, spectra compare in-phase surging motion ($\Delta\phi_S = 0$) and anti-phase motion ($\Delta\phi_S = \pi$) of the upstream turbine.
    }
    \label{fig:premult_compare}
\end{figure*}

In the absence of ambient turbulence, the relative phase between the surge motions of the two turbines, $\Delta \phi_S$, modifies the timing and spatial organization of the upstream wake but has a smaller influence on the overall wake recovery than the inflow turbulence scale. Figures~\ref{fig:SP_uniform}(a) and~\ref{fig:SP_uniform}(b) show the time-averaged streamwise velocity fields color-map for in-phase ($\Delta \phi_S = 0$) and anti-phase ($\Delta \phi_S = \pi$) motions under uniform and turbulent inflow conditions. For uniform inflow, both phase cases generate nearly identical wake shapes and velocity deficit levels, indicating that the phase difference alone does not produce strong reorganization of the coherent structures in the absence of ambient turbulence. The periodic surge motion does introduce a mild lateral oscillation in the near wake, but this effect remains largely symmetric and does not accumulate into significant changes by the downstream location at $x/D=5$.

When turbulence is introduced, and the integral length scale is large ($L_u/R = 1.25$), the effects of the rotors' phase difference become slightly more visible, although still secondary to the inflow-induced mixing. 
The turbulent eddies interact with the periodic platform motion and introduce additional unsteadiness in the shear layer, but the overall velocity deficit and wake width remain similar between the in-phase and anti-phase cases. This reinforces that wake development is primarily governed by the upstream turbine’s presence and the turbulence structure of the inflow, whereas the exact phase of surge motion plays only a minor role in shaping the mean wake. 

The pre-multiplied velocity spectra in Figure~\ref{fig:premult_compare} provide a further insight into the energy distribution. Under uniform inflow, Figure~\ref{fig:premult_compare}(a),  both phase cases exhibit similar spectral distributions at $x/D=6$, $7$, and $8$, with energy concentrated in comparable wavelength ranges. The phase difference shifts the amplitude of large-scale peaks slightly but does not change their location, confirming that the wake dynamics are dominated by the coherent structures shed from the upstream rotor rather than by the timing of the surge oscillation.

For turbulent inflow with the largest integral length scale ($L_u/R = 1.25$), the phase difference becomes more noticeable in the pre-multiplied spectra, Figure~\ref{fig:premult_compare}(b). At all downstream locations ($x/D = 6$--8), the anti-phase case ($\Delta\phi_S = \pi$) exhibits consistently higher energy peaks at the dominant wavelengths compared with the in-phase case. This indicates that anti-phase surge motion injects slightly stronger large-scale unsteadiness into the wake, which is then amplified by the highly coherent incoming turbulence. Nevertheless, despite the higher spectral amplitudes, the wavelength associated with the dominant wake structures remains nearly unchanged. The phase difference, therefore, alters the strength of wake meandering more than the underlying meander scale, and the resulting changes in the mean velocity deficit remain modest.

Overall, these results show that while the surging phase difference can influence the instantaneous wake structures and slightly enhance unsteadiness under turbulent inflow, its effect on the statistical wake properties is small. The wake dynamics and downstream recovery ability are governed primarily by the turbulence integral length scale, and the upstream turbine motion amplitude, with $\Delta\phi_S$ introducing only secondary modulations.

\subsection{Rotor Performance Analysis}

\begin{figure*}
  \centering
  \includegraphics[width=\textwidth]{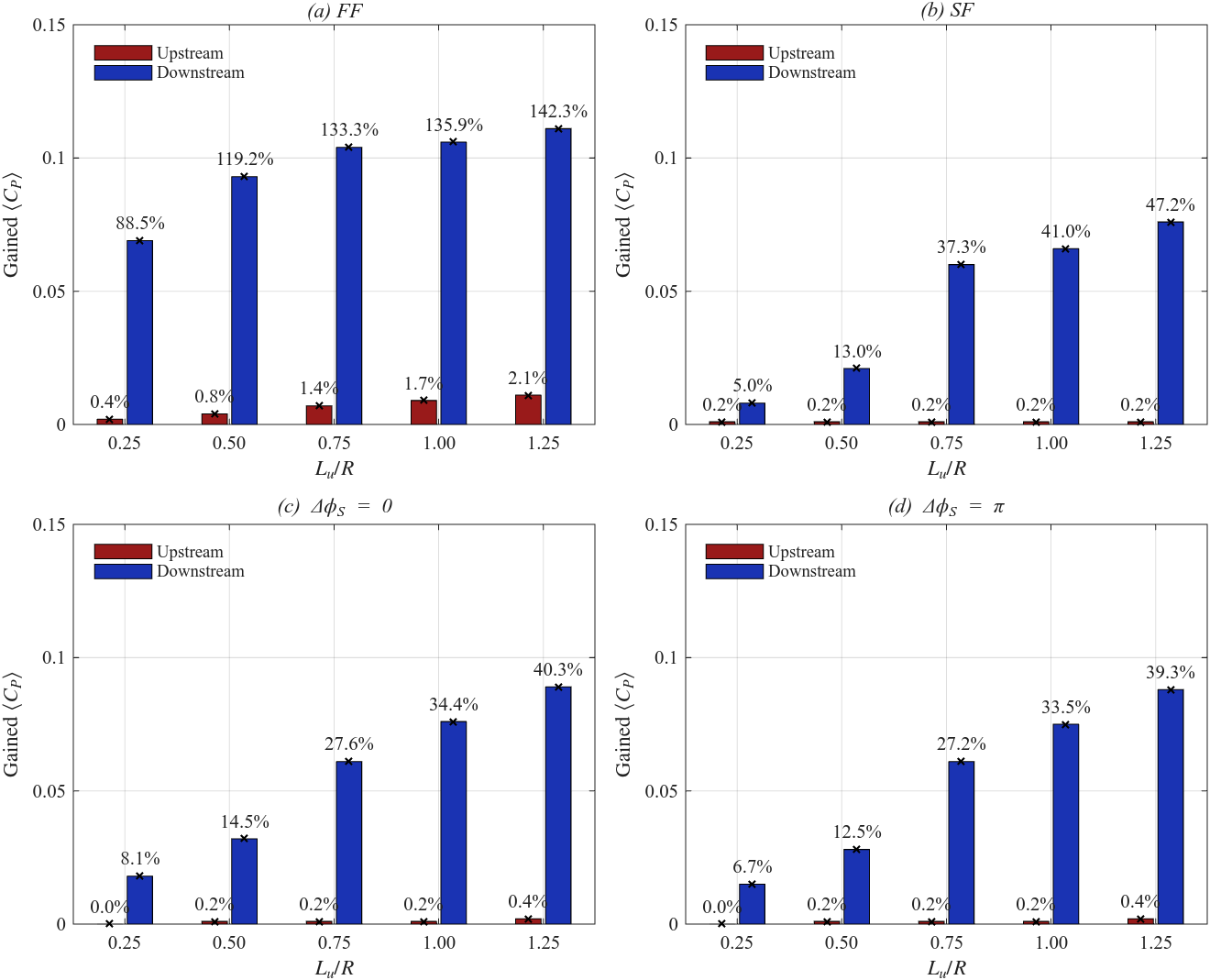}
  \caption{
    Comparison of upstream and downstream power gains relative to the uniform inflow baselines for various turbulence length scales $L_u/R$. 
    The four panels correspond to (a) fixed–fixed (FF), (b) surge–fixed (SF), (c) in-phase surging ($\Delta \phi_S=0$), and (d) anti-phase surging ($\Delta \phi_S=\pi$) configurations. 
    The blue bars represent downstream turbine gains, while red bars show upstream turbine variations, which remain nearly constant across all cases.
  }
  \label{fig:powerGained}
\end{figure*}

The impact of inflow turbulence scale and platform motion on turbine performance is now assessed through the time-averaged power coefficient, $\langle C_P \rangle$, with particular emphasis on the downstream rotor. The corresponding performance trends are illustrated in Figure~\ref{fig:powerGained}, where the power coefficient gain $\Delta C_P$ generated by each turbine is defined as the power coefficient obtained by each given turbine for the different inlet length scales studied $\langle C_P \rangle_{\text{turb}} $ versus the power coefficient generated by the corresponding turbine under uniform inflow conditions $\langle C_P \rangle_{\text{uniform}} $. 

\begin{equation}
\Delta C_P = \langle C_P \rangle_{\text{turb}} - \langle C_P \rangle_{\text{uniform}} 
\end{equation}

The corresponding output power coefficient values for the two turbines under uniform inflow conditions and for the different dynamic scenarios, which have served to obtain the power coefficient percentage increase introduced in Figure~\ref{fig:powerGained}, are defined in Table~\ref {tab:Cp_uniform}:

\begin{table}
\centering
\caption{Time-averaged power coefficients under uniform inflow for the upstream and downstream turbines in the four considered configurations.}
\label{tab:Cp_uniform}
\begin{ruledtabular}
\begin{tabular}{lcc}
Configuration & $\langle C_P^{up} \rangle$ & $\langle C_P^{down} \rangle$ \\
\hline
Fixed--Fixed (FF)            & 0.516 & 0.078 \\
Surge--Fixed (SF)            & 0.521 & 0.161 \\
Surge--Surge, in-phase ($\Delta \phi_S=0$)     & 0.521 & 0.221 \\
Surge--Surge, anti-phase ($\Delta \phi_S=\pi$) & 0.521 & 0.224 \\
\end{tabular}
\end{ruledtabular}
\end{table}

Across all configurations, the upstream turbine exhibits only a minor increase with the variations of the inflow turbulence length scale, confirming that changes in $\langle C_P^{up} \rangle$ remain small regardless of the movement of the upstream WT. 
In contrast, the downstream turbine shows a clear and monotonic increase in $\langle C_P^{down} \rangle$ with increasing $L_u/R$, as shown in Figure~\ref{fig:powerGained}. Depending on the configuration, the downstream turbine power gain reaches a considerable percentage relative to the corresponding uniform inflow baseline defined in Table ~\ref{tab:Cp_uniform}. 

Quantitatively, the magnitude of the downstream power gain depends strongly on the upstream turbine configuration. In the fixed–fixed (F–F) arrangement, increasing the turbulence length scale from $L_u/R = 0.25$ to $1.25$ leads to downstream power gains ranging from approximately $+88.5\%$ to $+142.3\%$ relative to the uniform inflow baseline. When the upstream turbine is allowed to surge while the downstream turbine remains fixed (S–F), the corresponding gains are more moderate, increasing from about $+5\%$ to $+47\%$ over the same range of $L_u/R$. These trends further confirm that the upstream turbine plays the dominant role in shaping the wake and governing downstream performance. For the cases in which both turbines are surging (S–S), whether in-phase or in anti-phase, only very minor differences in the gained downstream turbine power are observed, indicating that the relative phase of the downstream motion has a negligible influence on the time-averaged power response. Note that the power coefficient increase of the downstream turbine $\langle C_P^{down} \rangle$ is very similar for the SF, ($\Delta \phi_S=0$) and ($\Delta \phi_S=\pi$) configurations. 

We can conclude via saying that the influence of platform surge motion follows a similar but secondary pattern. Allowing the upstream turbine to surge further enhances wake unsteadiness and contributes to additional downstream power recovery, whereas variations in the relative phase of the surge motion between the two turbines produce only modest changes in the time-averaged power response (Figure~\ref{fig:powerGained}). This indicates that downstream performance is governed primarily by the upstream wake characteristics rather than by the motion state of the downstream platform itself.

These trends are fully consistent with the accelerated wake recovery observed for larger turbulence length scales. As demonstrated by the velocity deficit profiles, turbulence kinetic energy distributions, and Q-criterion visualizations, larger $L_u$ enhances low-frequency coherence, promotes earlier tip-vortex breakdown, and increases lateral and vertical entrainment. The resulting reduction in the persistence of the low-velocity wake core leads to a higher effective inflow velocity at the downstream rotor and, consequently, improved power extraction.


\section{Conclusions}

This study investigated how the integral length scale of inflow turbulence in a range of $0.25R \leq L_u \leq1.25R$, and platform surge motion influences wake development and power extraction in a tandem configuration of floating offshore wind turbines separated by five rotor diameters. Using large-eddy simulations with an actuator-line representation and synthetic turbulent inflow, the article evaluated the physical mechanisms through which inflow coherence and turbine motion, shape and modify wake structure, velocity recovery, and turbine–turbine interaction.

The results show that the inflow turbulence integral length scale plays a dominant role in governing wake dynamics. Over the range considered, increasing $L_u$ is accompanied by an increase in the free-stream $TI$ from approximately $1.9\%$ to $7.2\%$, reflecting a stronger coupling between these two turbulence parameters. Large integral scales introduce energetic, low-frequency eddies that interact strongly with the rotor-generated shear layers, leading to earlier destabilization of the tip-vortex system, enhancing lateral and vertical entrainment, and accelerating wake recovery. These effects manifest consistently across velocity deficit profiles, turbulent kinetic energy distributions, spectral signatures, and vortex topology, indicating a robust transition from a coherent, vortex-dominated near wake to a rapidly mixing turbulent wake as the inflow scale increases.

Superimposed platform surge motion of the upstream turbine further promotes wake unsteadiness by periodically perturbing the tip-shear layer and exciting flow momentum interchange. 
This mechanism reinforces inflow-driven mixing and leads to additional reductions in the inter-turbine velocity deficit. However, once strong background turbulence is present, the relative contribution of surge motion becomes secondary, and wake recovery is primarily controlled by the flow turbulence length scale and its associated intensity. 
When both turbines are allowed to surge, the relative phase of their motion modifies instantaneous wake features but has only a minor impact on the mean wake and time-averaged power response.

The wake-physics trends translate directly into turbine performance. The downstream power gain increases monotonically with inflow turbulence integral length scale and is strongly influenced by the upstream turbine configuration, while the upstream turbine itself remains largely insensitive to changes in inflow scale. These findings demonstrate that downstream performance in floating wind turbine arrays is governed primarily by upstream wake dynamics rather than by the motion state of the downstream platform.

Taken together, this work provides a physics-based understanding of turbulence–motion coupling in floating wind turbine wakes. It highlights the central role of inflow turbulence coherence in shaping wake recovery and clarifies how platform motion modifies, but does not fundamentally control, this process. These insights are essential for interpreting wake interactions in realistic offshore environments and for developing modeling and control strategies that account for both atmospheric turbulence structure and platform dynamics.

It should be noted that in all turbulent inflow simulations considered in this study, the Reynolds stress level imposed at the inlet was kept constant, while only the integral length scale field was varied. From the results obtained, it was realized that changes in the turbulence length scale at the computational domain inlet were accompanied by variations in the resulting turbulence intensity when measured further downstream, as shown in Figure~\ref{fig:ti-inflow}. The present approach reflects realistic coupling between turbulence scale and intensity, as can be extracted from equations~\ref{turb_intensity} to \ref{two_point_correlation}, but it does not allow their effects to be fully isolated. Future work could therefore explore inflow configurations 
to more clearly evaluate the respective roles of turbulence strength and coherence on wake dynamics and turbine interaction.

\begin{acknowledgments}

The project received financial support from the Spanish Ministry of Science, Innovation, and Universities through the project PID2023-150014OB-C21. Access to high-performance computing facilities was provided by the Spanish Supercomputing Network (Red Española de Supercomputación, RES) under allocations IM-2024-3-0003 and IM-2025-1-0004. The first author is funded by the 2025-DI-00014 Industrial PhD grant, awarded by the Generalitat de Catalunya with the support of the Y-Plasma Enterprise.
We would like to sincerely thank  Dr. Navid M. Tousi, Dr. Emily Louise Hodgson, Dr. Søren Andersen, and Mr. YuanTso Li for the helpful discussions and comments we had during the development of this paper.
 
\end{acknowledgments}

\section*{Author DECLARATION}

\textbf{Ahmad Nabhani}: Conceptualization (equal); Methodology (equal); Software (equal); Investigation (lead); Formal analysis (lead); Visualization (lead); Writing – original draft (lead); Writing – review \& editing (equal). 
\textbf{Josep M Bergadà}: Supervision (lead); Conceptualization (equal); Methodology (equal); Software (equal); Investigation (supporting); Formal analysis (supporting); Writing – original draft (supporting); Writing – review \& editing (equal), Project manager.

\section*{Data Availability Statement}

The data that support the findings in this study are available from the corresponding author upon reasonable request.

\appendix

\section{ Comparison of the present study with previous research}
\label{app1}

Table \ref{tab:NREL5MW} introduces the comparison of the time-averaged power and thrust coefficients obtained from the present study with some previous works. The table further demonstrates the good reliability associated with the present research, since the results agree very well with the previous ones, particularly with the very recent research undertaken by \cite{li2024wake}.

\begin{table*}
\centering
\caption{Comparison of the thrust coefficients $\langle C_T \rangle$ and power coefficients $\langle C_P \rangle$ from our study of a fixed NREL 5MW rotor under its rated conditions with results from other previous numerical studies.}
\label{tab:NREL5MW}
\begin{ruledtabular}
\begin{tabular}{lcccccc}
\textbf{Source} & \textbf{Turbulence model} & \textbf{Force model} & \textbf{TI [\%]} & \textbf{Configuration} & $\langle C_T \rangle$ & $\langle C_P \rangle$ \\
\hline
Present & LES  & ALM & Uniform & Fix  & 0.736 & 0.516 \\
Present & LES  & ALM & Uniform & Surg & 0.729 & 0.521 \\
\hline
Li et al.~\cite{li2024wake} & LES  & ALM & Uniform & Fix  & 0.728 & 0.518 \\
Li et al.~\cite{li2024wake} & LES  & ALM & 5.3     & Fix  & 0.727 & 0.518 \\
Jonkman et al.~\cite{Jonkman2009} & ---  & BEM & --- & Fix & 0.81  & 0.47  \\
Johlas et al.~\cite{Johlas_2019} & LES  & ALM & 4.1 & Surg & 0.744 & ---   \\
Xue et al.~\cite{en15010282} & LES  & ALM & Uniform & Fix & 0.75  & 0.52  \\
Li et al.~\cite{li2015numerical} & RANS & ALM & --- & Surg & 0.712 & 0.451 \\
Yu et al.~\cite{app8030434} & RANS & ALM & --- & Fix  & 0.728 & 0.472 \\
Rezaeiha et al.~\cite{REZAEIHA2021859} & RANS & ADM & 5 & Surg & 0.715 & 0.567 \\
\end{tabular}
\end{ruledtabular}
\end{table*}

\nocite{*}
\bibliography{aipsamp}

\end{document}